\documentclass[preprint,prd,aps,showpacs,groupedaddress]{revtex4-1}
\usepackage{amsmath,hyperref,amssymb,amsfonts,amsthm,dcolumn,color,graphicx,graphics,latexsym,placeins,epsfig}
\usepackage{subfigure,float,psfrag,rotating}
\usepackage{natbib}
\hypersetup{colorlinks=true,citecolor=blue,linkcolor=blue,urlcolor=blue}

\begin{document}

\title{Born-Infeld gravity with a Brans-Dicke scalar}

%\author{Soumya Jana and Sayan Kar}
%\email{soumyajana@phy.iitkgp.ernet.in, sayan@phy.iitkgp.ernet.in}
%\affiliation{\rm Department of Physics {\it and} Centre for Theoretical Studies \\Indian Institute of Technology, Kharagpur, 721302, India}

\author{Soumya Jana}
\email{soumyajana@phy.iitkgp.ernet.in}

\author{Sayan Kar}%
 \email{sayan@phy.iitkgp.ernet.in}
\affiliation{\rm Department of Physics {\it and} Centre for Theoretical Studies \\Indian Institute of Technology, Kharagpur, 721302, India}

\begin{abstract}
Recently proposed  Born-Infeld (BI) theories of gravity 
assume a constant BI parameter ($\kappa$). However, no clear 
consensus exists on the sign and value of 
$\kappa$. Recalling the Brans-Dicke (BD) approach, where a scalar field was 
used to generate the gravitational constant $G$, we suggest
an extension of Born-Infeld gravity with a similar Brans-Dicke flavor.
Thus, a new action, with $\kappa$ elevated to a spacetime dependent real 
scalar field, is proposed. 
We illustrate this new theory in a cosmological setting with pressureless 
dust and radiation as matter. Assuming a functional form of $\kappa(t)$, 
we numerically obtain the scale factor evolution and other details of the 
background cosmology. It is known that BI gravity differs from general relativity (GR) in the 
strong-field regime but reduces to GR for intermediate and weak fields. 
Our studies in cosmology demonstrate how, with this new, scalar-tensor 
BI gravity, deviations from GR as well as usual BI gravity, 
may arise in the weak-field regime too. For example, we note a late-time 
acceleration without any dark energy contribution. Apart from such
qualitative differences, we note that
fixing the sign and value of $\kappa$ is no longer a necessity in this theory,
though the origin of the BD scalar does remain an open question.
\end{abstract}


\maketitle

\section{Introduction} 
\label{sec:1}
\noindent General relativity (GR) is surely successful as a classical theory 
of gravity, and more so, with the recent detection of gravitational waves \cite{abbott}. Over the years, it has passed through several precision tests without any significant sign of failure. However, most of these tests \cite{Will2014} are either in vacuum or in the weak-field regime. They largely verify 
the Einstein equivalence principle and set constraints on 
weak-field deviations from GR, as encoded through the parametrized post-Newtonian formalism. 
 
\noindent On the other hand, the occurrence of spacetime singularities 
under very reasonable assumptions on causal structure 
and matter stress-energy has been shown many years ago in the work of
Hawking and Penrose \cite{hawk,wald}. 
Singularities (cosmological, black hole or naked) are thus
unavoidable. Therefore, a resolution of singularities and/or an 
understanding about the consequences of their existence is highly desirable.

\noindent It is also a fact that, despite 
immense theoretical efforts, an explanation of the origin of dark matter or 
dark energy does not seem to exist within the framework of GR. 
The need of an understanding/solution 
to the dark matter and dark energy problems
stem from the fact that both of them arose from observations. 
For recent reviews on dark energy and dark matter see \cite{copeland,freese}.

\noindent In order to address some of these problems, 
it is not unusual to construct classical 
theories which deviate from GR, particularly in the strong-field regime. 
Thus, we have various proposals on modified gravity \cite{mod_grav_rev2,mod_grav_rev1} 
at the classical level,
apart from the intense pursuit of quantum gravity \cite{bojowald2001,ashtekar2006}.
A modified gravity model must necessarily have a gravitational action
which is different from the standard Einstein-Hilbert action. It is
also true that there are, within GR, several models (particularly
for dark energy \cite{copeland}) which assume various types of rather nonstandard
matter stress-energy. We will, however focus here  on modifications in
the gravity sector only.

\noindent One such modified gravity model is inspired by
Born-Infeld (BI) electrodynamics where 
the infinity in the electric field at the location of a 
point charge is regularized \cite{born}. 
With a similar determinantal structure 
$\left(\left[\sqrt{-det(g_{\mu\nu}+\kappa R_{\mu\nu})} \right]\right)$ in the 
action, a gravity theory in the metric formulation was first suggested by 
Deser and Gibbons \cite{desgib}. 
In fact, a determinantal form of the gravity action existed 
in Eddington's affine reformulation of GR for de Sitter spacetime 
\cite{edd}, though matter coupling remained a problem 
in the Eddington approach.

\noindent Much later, Vollick \cite{vollick} introduced the Palatini 
formulation of  Born-Infeld gravity and worked on various related aspects. 
Unlike metric variation, where the connection is assumed to be
related to the metric, in a Palatini variation, both the metric
and connection are varied independently. Consequences of
these two approaches regarding the existence of additional propagating 
degrees of freedom (in the metric approach), absent in the Palatini
formulation, 
as well as a general review on the Palatini approach in modified gravity
can be found in \cite{olmo2011}. 

\noindent Vollick also introduced a nontrivial and somewhat artificial way 
of coupling matter in his theory \cite{vollick2,vollick3}. More recently, 
Ba\~nados and Ferreira  \cite{banados} have come up with a formulation 
where matter coupling is 
different and simpler compared to Vollick's proposal. 
We focus here on the theory proposed in Ref. \cite{banados} and refer to it
as Eddington-inspired Born--Infeld (EiBI) gravity, for obvious reasons. 
The EiBI theory reduces to GR in vacuum.
It also falls within the class of bimetric 
theories of gravity (bigravity) \cite{isham}, \cite{scargill}, 
\cite{schmidt,lavinia}.   

\noindent Let us first briefly recall Eddington--inspired Born--Infeld (EiBI)
gravity. The central feature here is the existence of a physical metric 
which couples to matter and another auxiliary metric which is not used for 
matter couplings. 
One needs to solve for both metrics through the field equations. The action for the theory developed in Ref. \cite{banados} is given as
\begin{equation}
S_{BI}(g,\Gamma, \Psi) =\frac{c^3}{8\pi G\kappa}\int d^4 x \left [ \sqrt{-\vert g_{\mu\nu} +\kappa R_{\mu\nu}(\Gamma)\vert}-\lambda \sqrt{-g} \right]+ S_M (g, \Psi),
\end{equation}
where $\lambda=\kappa \Lambda +1$, with $\Lambda$ being the cosmological constant. A Palatini variation with respect to $g_{\mu\nu}$ and $\Gamma$, using the
auxiliary metric $q_{\mu\nu}=g_{\mu\nu}+\kappa R_{\mu\nu}(\Gamma)$ and assuming $R_{\mu\nu}$ symmetric, gives the field equations for this theory.

%\begin{equation}
%q_{\mu \nu}=g_{\mu\nu} + \kappa R_{\mu\nu}(q).
%\label{eq:gammavarn}
%\end{equation}
%Variation with respect to $g_{\mu\nu}$ gives 
%\begin{equation}
%\sqrt{-q} q^{\mu\nu} = \lambda \sqrt{-g}g^{\mu\nu}-\frac{8\pi G}{c^4}\kappa \sqrt{-g} T^{\mu\nu},
%\label{eq:gvarn}
%\end{equation}
%where $T^{\mu\nu}$ are components of stress-energy tensor in the coordinate frame.

\noindent In order to obtain solutions, we need to assume a $g_{\mu\nu}$ and 
a $q_{\mu\nu}$
with unknown functions, as well as a matter stress-energy ($T^{\mu\nu}$). 
Thereafter, we write down the field equations and obtain solutions using 
some additional assumptions about the metric functions and the stress-energy. 

\noindent A lot of work on various fronts has been carried out
on diverse aspects of this theory, in the last few years.
Astrophysical scenarios have been widely discussed \cite{cardoso,casanellas,avelino,sham,sham2,structure.exotic.star,sotani.neutron.star,
sotani.stellar.oscillations,sotani.magnetic.star}. Spherically symmetric 
solutions of various types have been obtained 
\cite{banados,wei,sotani,eibiwormhole,rajibul,jana2,BTZ_typesoln,scalar_geon_BIgravity}.
A domain wall brane in a higher-dimensional generalization of EiBI theory
was analyzed in Ref. \cite{eibibrane}. 
Generic features of the paradigm of matter-gravity couplings were analyzed in 
\cite{delsate}. Further, in  \cite{cho_prd88}, the authors showed that 
EiBI theory admits a nongravitating matter distribution, which is not allowed 
in GR. Some interesting cosmological and circularly symmetric solutions in $2+1$ dimensions are obtained in \cite{jana}. In \cite{pani}, a problem
in the context of stellar physics, related to surface singularities in 
EiBI gravity, was noticed. Gravitational 
backreaction was suggested as a cure in \cite{eibiprob.cure}. 
A modification of EiBI theory, through a functional extension similar to 
$f(R)$ theory, was proposed in \cite{odintsov}.  
Recently, in \cite{fernandes} a new route to  matter 
coupling was suggested via the use of the Kaluza ansatz in a 
five-dimensional EiBI action (in a metric formulation) and 
subsequent compactification to four-dimensional gravity coupled 
nonlinearly to electromagnetism. Generalization of the EiBI theory by adding a pure trace term in the determinantal action was suggested in \cite{chen2016} and some interesting cosmological solutions were found, such as a de-Sitter stage in a radiation dominated Universe.

\noindent A lot of the recent work on EiBI gravity is devoted to cosmology. 
In \cite{banados,scargill,cho}, the nonsingularity 
of the Universe filled by any ordinary matter was demonstrated. 
Linear perturbations have 
been studied in the background of homogeneous and isotropic spacetimes 
in the Eddington regime \cite{escamilla,linear.perturbation}. Bouncing 
cosmology in EiBI gravity was emphasized as an alternative to  inflation 
in \cite{avelinoferreira}. The authors in \cite{chokim}, studied a 
model described by a scalar field with a quadratic potential, which results 
in a nonsingular initial state of the Universe leading naturally to inflation. 
They also investigated the stability of tensor perturbations in this 
inflationary model \cite{cho90}, whereas the scalar perturbations were studied 
in \cite{cho_scalar_perturbation}. 
Large-scale structure formation in the Universe and the integrated 
Sachs-Wolfe effect were discussed in \cite{large.scale.structure}. 
Quantum effects near the late-time abrupt events was studied in the EiBI model by proposing an effective Wheeler-DeWitt equation \cite{Bouhmadi-Lopez2016,lopez2017} and it was shown that these events are expected to be avoided when quantum effects are under consideration.
Other relevant work has been reported in  \cite{cho_spectral_indices,cho_pow_spectra,cho_tensor_scalar,felice,power.spectrum,
cascading_dust_inflation,bianchi.cosmo,jana4,instanton,Bouhmadi-Lopez2014,Bouhmadi-Lopez2015,lopez2014}. For a very recent review on Born-Infeld gravity, 
see \cite{jimenez2017} and the references therein.

\noindent The theory parameter $\kappa$ in EiBI gravity is a constant
though we have no way to know whether it is universal. The sign of $\kappa$ governs the nature of
solutions and its value determines the scale at which corrections to GR dynamics cannot be neglected. There are some upper bounds on the value of $\kappa$ from 
astrophysical and cosmological observations \cite{cardoso,casanellas,avelino,nuclear.test}. For example, the existence of self-gravitating compact objects like neutron stars strongly constrains the theory with $\kappa > 0$ and $\kappa \lesssim 5\times 10^8 $ m$^2$ \cite{cardoso}. Stellar equilibrium and evolution of 
the Sun puts a constraint $\vert\kappa\vert \lesssim 2\times 10^{14}$ m$^2$ \cite{casanellas}. Primordial nucleosynthesis leads to $\kappa \lesssim 10^6$ m$^2$ \cite{avelino} where it is assumed that $\kappa>0$. From nuclear physics 
constraints (i.e. requiring the electromagnetic force as dominant 
over the gravitational force, at the subatomic scale) one gets 
$\vert \kappa \vert \lesssim 6\times 10^5$ m$^2$ \cite{nuclear.test}. All the 
numbers (for $\kappa$) mentioned above are in the unit of m$^2$, whereas, in 
most of the literature, the unit used (for $\kappa'= 8\pi G \kappa$) is kg$^{-1}$m$^5$s$^{-2}$. In summary, no consensus exists on the sign and value of 
$\kappa$. 

\noindent In our work here, we address this issue by suggesting 
the possibility of $\kappa$ being a nonconstant, real scalar field. 
The advantage with $\kappa$ being a scalar field is that it can take on
different functional forms in different scenarios (say, cosmology, black holes, stars etc.) and a universal sign or value is not a necessity. However, one still
needs to address the issue of the origin of $\kappa$. 

\noindent It is known that EiBI theory differs from GR in the high energy 
regime. With a scalar $\kappa$ a new theory of gravity emerges, which reduces 
to GR only in the intermediate energy scale, but may differ in the high as 
well as the low energy regimes. Our aim here is to formulate this theory 
with a scalar $\kappa$ and explore its consequences. This is carried out in the
subsequent sections.

\section{The EiBI action with $\kappa$ as a real scalar field}
\label{sec:action}

Let us begin by proposing a new action given as 
\begin{eqnarray}
S_{BI\kappa}\left(g,\Gamma,\kappa,\Psi\right)&=& \int \left[\frac{1}{\kappa}\left(\sqrt{-\vert g_{\alpha\beta}+\kappa R_{\alpha\beta}(\Gamma)\vert} -\sqrt{-g} \right)-\sqrt{-g}\tilde{\omega} (\kappa)g^{\mu\nu}\partial_{\mu}\kappa \partial_{\nu}\kappa \right]d^Dx \nonumber\\ 
&& + S_M\left(g,\Psi\right),
\label{eq:modified_action}
\end{eqnarray}
where $\kappa(t,\vec{x})$  is a scalar field and $\tilde{\omega}(\kappa)$ is 
a coupling function, reminiscent of scalar-tensor (Brans-Dicke) 
modifications of GR \cite{faraoni}.  
We assume $c=1$, $8\pi G=1$ and spacetime of dimension  $D$. We also assume the Ricci tensor ($R_{\alpha\beta}$) to be symmetric.
For a constant $\kappa$ we recover the standard EiBI theory of gravity
\cite{banados}. 
If $\kappa$ is constant and small in value, the action reduces to the known 
Einstein-Hilbert one (with cosmological constant $\Lambda =0 $). 
Variation of the action [Eq.~(\ref{eq:modified_action})] with respect to `$\Gamma$' yields the earlier definition of the auxiliary metric field,
\begin{equation}
q_{\alpha\beta}=g_{\alpha\beta}+\kappa R_{\alpha\beta}(q),
\label{eq:gammavarn_modify}
\end{equation} 
where the $\Gamma$s are computed using the following relation
\begin{equation}
\Gamma^{\alpha}_{\mu\nu}=\frac{1}{2}q^{\alpha\beta}\left(\partial_{\nu}q_{\beta\mu} +\partial_{\mu} q_{\nu\beta}-\partial_{\beta}q_{\mu\nu} \right),
\label{eq:gamma}
\end{equation}
as the connection satisfies the standard metric-connection compatibility with the metric $q_{\mu\nu}$, i.e. $\tilde{\nabla}_{\mu}\left(\sqrt{-q}q^{\alpha\beta}\right)=0$.
However variation with respect to `$g_{\alpha\beta}$' yields
\begin{equation}
\sqrt{-q}q^{\alpha \beta}-\sqrt{-g}g^{\alpha\beta}=-\kappa\sqrt{-g}\,T^{\alpha \beta}_{eff},
\label{eq:gvarn_modified}
\end{equation}
where
\begin{equation}
T^{\alpha \beta}_{eff}=T^{\alpha \beta}-\tilde{\omega} g^{\alpha\beta}g^{\mu\nu}\partial_{\mu}\kappa\partial_{\nu}\kappa +2\tilde{\omega} g^{\mu\alpha}g^{\nu\beta}\partial_{\mu}\kappa\partial_{\nu}\kappa .
\label{eq:tmunu_eff}
\end{equation}
$T^{\alpha\beta}$ is the usual stress-energy tensor. Variation with respect 
to $\kappa$ gives 
\begin{equation}
2\kappa \tilde{\omega }(\kappa)\nabla_{\mu}\nabla^{\mu}\kappa+\kappa \tilde{\omega} '(\kappa)\nabla_{\mu}\kappa \nabla^{\mu}\kappa +\frac{1}{\kappa} + \frac{\sqrt{-q}}{\sqrt{-g}} \left(\frac{1}{2}q^{\alpha \beta}R_{\alpha \beta}(q)-\frac{1}{\kappa}\right)=0,
\label{eq:kappavarn}
\end{equation}
where the covariant derivatives are defined with respect to the physical metric ($g$) and $\tilde{\omega}'(\kappa)$ is a derivative of $\tilde{\omega}$ with respect to $\kappa$.

\noindent Using the abovementioned field equations, one can verify that the 
stress-energy tensor ($T^{\mu\nu}$) is conserved, i.e.
\begin{equation}
\nabla_{\mu}T^{\mu\nu}=0.
\end{equation} 

\noindent It is important to check whether the above equations are consistent 
with the solutions for constant $\kappa$ --particularly Eq.~(\ref{eq:kappavarn}). In vacuum, from Eq.~(\ref{eq:gvarn_modified}), we have $\sqrt{-q}q^{\alpha \beta}=\sqrt{-g}g^{\alpha\beta}$ which implies $q_{\mu\nu}=g_{\mu\nu}$. Using this in Eq.~(\ref{eq:gammavarn_modify}), $R_{\alpha\beta}=0$. Hence, Eq.~(\ref{eq:kappavarn}) is satisfied. Now, to check the consistency in presence of a
 matter distribution ($T_{\alpha\beta}\neq0$), we take the example of a 
three-dimensional ($D=3$) cosmological solution in EiBI gravity for a 
dust-filled ($P=0$) Universe \cite{jana}. The physical Friedmann-Robertson-Walker (FRW) spacetime is given by $ds^2=-dt^2+a^2(t)[dr^2+r^2d\theta^2]$, where $a^2(t)=\rho_0 (t^2-\kappa)$ for $\kappa>0$ and $\kappa<0$ as well, and $\rho_0$ is the present day energy density of the Universe. The corresponding auxiliary line element is $ds_q^2=-dt^2+b^2(t)[dr^2+r^2d\theta^2]$, where $b^2(t)=\rho_0 t^2$. Then, $R(q)=2\left(\frac{\dot{b}^2}{b^2}+2\frac{\ddot{b}}{b}\right)=2/t^2$. Using these relations, it is now easy to verify that Eq.~(\ref{eq:kappavarn}) is consistent for a constant $\kappa$.

\noindent The nonrelativistic limit of the theory is different from that in 
EiBI gravity \cite{banados}. For a time-independent physical metric 
$ds^2=-(1+2\Phi)dt^2+(1-2\Phi)d\vec{x}\cdot d\vec{x}$ and an energy-momentum 
tensor $T^{\mu\nu}=\rho u^{\mu}u^{\nu}$, the full set of linearized field 
equations are given by the following two equations:
\begin{eqnarray}
\nabla^2\Phi=\frac{\rho}{2}+\frac{1}{4}\nabla^2(\kappa \rho)+\frac{1}{2}\tilde{\omega}(\vec{\nabla}\kappa)^2+\frac{1}{4}\nabla^2\left(\kappa \tilde{\omega}(\vec{\nabla}\kappa)^2\right),\label{eq:poisson_modfiedeibi}\\
2\tilde{\omega}\nabla^2\kappa +\tilde{\omega}'(\vec{\nabla}\kappa)^2+\frac{1}{4}\left(\rho+\tilde{\omega}(\vec{\nabla}\kappa)^2\right)^2=0,
\end{eqnarray} 
where $\Phi$, $\rho$, and $\kappa$ depend only on $\vec{x}$. Equation~(\ref{eq:poisson_modfiedeibi}) is the modified Poisson equation in the new theory. 
For a constant $\kappa$ it reduces to the Poisson equation in the
original EiBI theory. 

\noindent We also mention that a study of gravitational waves in vacuum
as well as vacuum exact solutions
in this theory
will be different (unlike standard EiBI gravity \cite{banados}) 
from usual GR  because of the presence of the scalar 
field $\kappa$.

\section{Cosmology}

\noindent As an application of the new theory, we now 
study cosmology in the ($3+1$)-dimensional version of the new theory. 
We assume a spatially flat, FRW ansatz 
for the physical line element:
\begin{equation}
ds^2=-dt^2+a^2(t)\left[dx^2+dy^2+dz^2\right],
\label{eq:gmetric_ch6}
\end{equation}
and choose an ansatz for the auxiliary line element
\begin{equation}
ds^2_q=-Udt^2+Va^2\left[dx^2+dy^2+dz^2\right].
\label{eq:qmetric_ch6}
\end{equation}
Let us consider a Universe driven by a perfect fluid with the 
stress-energy tensor,
\begin{equation}
T^{\mu\nu}=(p+\rho)u^{\mu}u^{\nu}+pg^{\mu\nu},
\end{equation}
where $p$ and $\rho$ are pressure and energy density respectively, 
and $u^{\mu}=diag.\lbrace 1,0,0,0\rbrace$. 
Using Eqs.~(\ref{eq:gmetric_ch6}) and (\ref{eq:qmetric_ch6}), the `00' (temporal) and `$ii$' (spatial; $i=1,2,3$) components of $T^{\mu\nu}_{eff}$ [Eq.~(\ref{eq:tmunu_eff})] become, 
\begin{equation}
T^{00}_{eff}=\rho +\tilde{\omega} \dot{\kappa}^2 ~\quad \mbox{and}~\quad  T^{ii}_{eff}= (p +\tilde{\omega} \dot{\kappa}^2)/a^2.
\end{equation}

\noindent Further use of Eq.~(\ref{eq:gvarn_modified}) leads to expressions 
for $U$ and $V$ given by
\begin{eqnarray}
U&=&\frac{\left(2-y-\kappa \omega \rho\right)^{3/2}}{\sqrt{y+\kappa \rho}},\label{eq:U_ch6}\\
V&=&\sqrt{(y+\kappa \rho)(2-y-\kappa \omega \rho)}\, ,
\label{eq:V_ch6}
\end{eqnarray}
where we have defined a new variable $y=1+\kappa \tilde{\omega}\dot{\kappa}^2$ 
and used the equation of state $p=\omega \rho$, with $\omega$ being a constant. 
The `00' and `11' equations resulting from  `$\Gamma$'-variation lead to
\begin{eqnarray}
\frac{\ddot{a}}{a}+\frac{\ddot{V}}{2V}-\frac{\dot{V}^2}{4V^2}+\frac{\dot{a}}{a}\frac{\dot{V}}{V}-\frac{\dot{U}}{2U}\left(\frac{\dot{a}}{a}+\frac{\dot{V}}{2V}\right)=\frac{U-1}{3\kappa},
\label{eq:R_00(q)_ch6}\\
\frac{\ddot{a}}{a}+\frac{\ddot{V}}{2V}+\frac{\dot{V}^2}{4V^2}+3\frac{\dot{a}}{a}\frac{\dot{V}}{V}-\frac{\dot{U}}{2U}\left(\frac{\dot{a}}{a}+\frac{\dot{V}}{2V}\right)+2\left(\frac{\dot{a}}{a}\right)^2=\frac{1}{\kappa}\left(U-\frac{U}{V}\right).
\label{eq:R_11(q)_ch6}
\end{eqnarray} 

\noindent Subtracting Eq.~(\ref{eq:R_00(q)_ch6}) from Eq.~(\ref{eq:R_11(q)_ch6}) 
we obtain
\begin{equation}
\left(\frac{\dot{a}}{a}+\frac{\dot{V}}{2V}\right)^2=\frac{1}{6\kappa}\left(1+2U-3\frac{U}{V}\right).\label{eq:R_eq}
\end{equation}

\noindent The $\kappa$-variation equation (\ref{eq:kappavarn}) becomes
\begin{equation}
\dot{y}+6(y-1)\frac{\dot{a}}{a}=\dot{\kappa}\left[\frac{1}{2\kappa}\left(y+\kappa \rho\right)\left(1+2U-3\frac{U}{V}\right)-\rho\right].
\label{eq:kappavarn_cosmo}
\end{equation}

\noindent Finally, conservation of the stress-energy tensor leads to
\begin{equation}
\dot{\rho}+3(\omega +1)\rho \frac{\dot{a}}{a}=0,
\label{eq:conservationeq_ch6}
\end{equation} 
which yields the same GR relation between $\rho$ and $a$.

\noindent Thus, we have five independent equations [Eqs.~(\ref{eq:U_ch6}), (\ref{eq:V_ch6}), (\ref{eq:R_eq}), (\ref{eq:kappavarn_cosmo}), and (\ref{eq:conservationeq_ch6})] to solve for six unknown functions ($a$, $U$, $V$, $\kappa$, $\rho$, and $y$). Hence we have the freedom to choose a form of $\kappa(t)$, which 
we assume as
\begin{equation}
\kappa(t)=\kappa_0+\epsilon \exp(\mu t),
\label{eq:kappa_t}
\end{equation}
with  $\kappa_0$, $\epsilon$, and $\mu$ as constants. 
For $\mu>0$, $\kappa\rightarrow \kappa_0$ at $t\rightarrow -\infty$ and, for $\mu<0$, $\kappa\rightarrow \kappa_0$ at $t\rightarrow \infty$. In limiting situations, where $\vert\kappa_0\vert\gg \vert \epsilon \exp(\mu t) \vert$, we expect to recover the known EiBI gravity (for a constant $\kappa$) and, in the other
regime, there may be deviations from EiBI gravity. In the following 
subsections, we investigate possible deviations for the three cases: 
(i) vacuum, (ii) dust ($p = 0$), and (ii) radiation ($p= \rho/3$). 

\subsection{Vacuum }
\begin{figure}[!htbp]
\centering
\subfigure[]{
\mbox{\epsfig{file=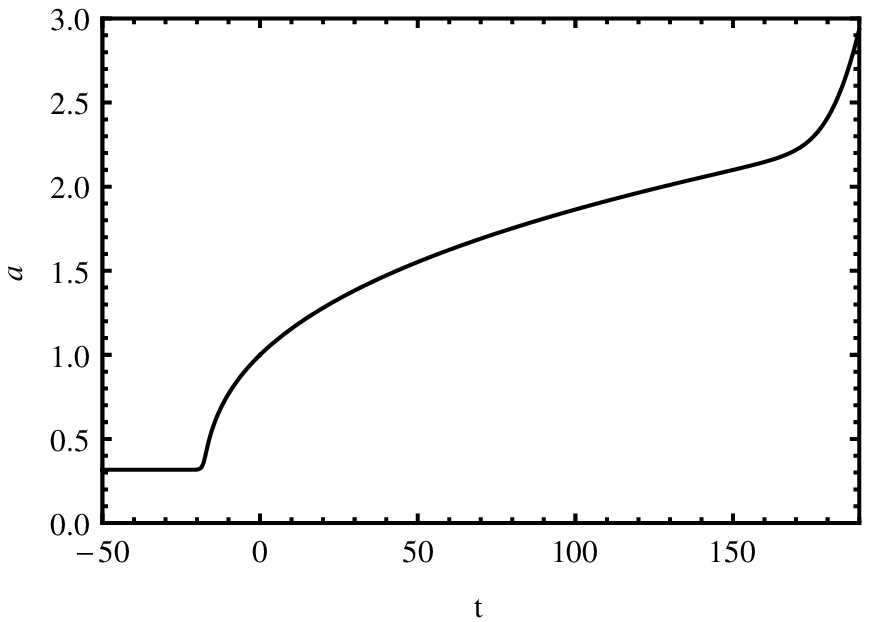,width=0.45\textwidth,angle=360}}
\label{fig:a_vac}
}
\subfigure[]
{
\mbox{\epsfig{file=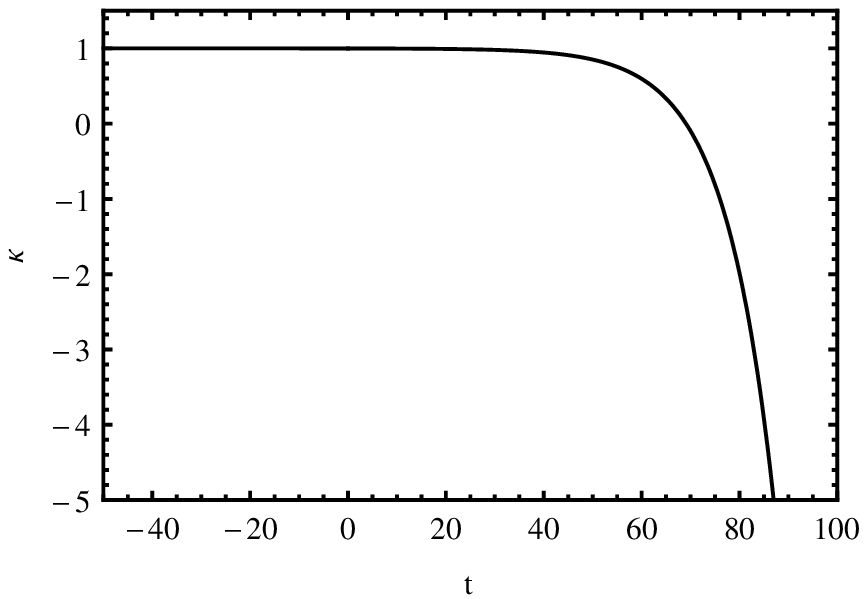,width=0.45\textwidth,angle=360}}
\label{fig:k_vac}
}\\
\subfigure[]
{
\mbox{\epsfig{file=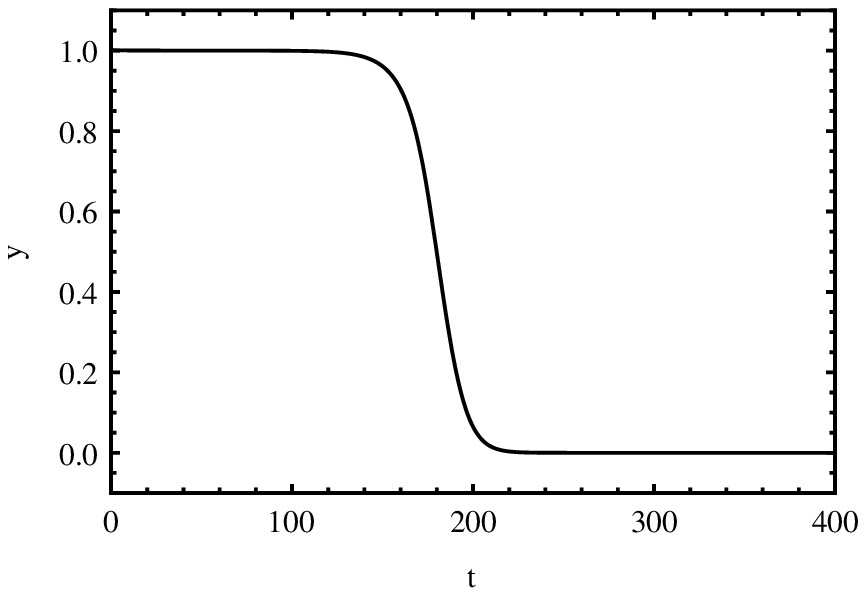,width=0.45\textwidth,angle=360}}
\label{fig:y_vac}
}
\subfigure[]{
\mbox{\epsfig{file=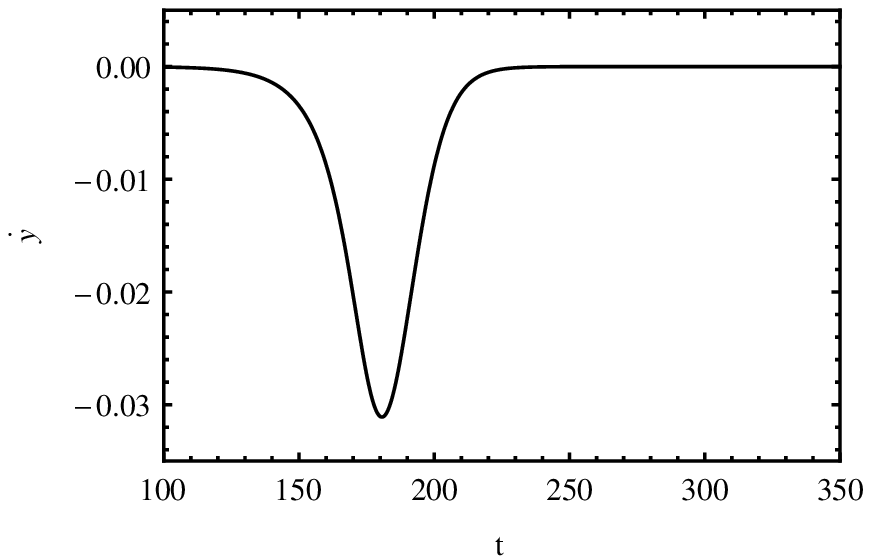,width=0.45\textwidth,angle=360}}
\label{fig:doty_vac}}
\caption{Plot of (a) $a(t)$, (b) $\kappa(t)$, (c) $y(t)$, and (d) $\dot{y}(t)$ for a vacuum ($\rho=0$ and $p=0$) solution for $\kappa_0 >0$, $\epsilon<0$, and $\mu>0$. The parameters used are $\kappa_0=1$, $\mu=0.1$. We choose $a(0)=1$, $y(0)=1.001$, $\kappa(0)=0.999$ for the numerical solution. }
\label{fig:ak_vac}
\end{figure}

\noindent Unlike GR or standard EiBI gravity, in this new theory, we do have nontrivial vacuum FRW solutions generated primarily by the time-dependent scalar field $\kappa(t)$. For $\mu>0$ (see Eq.~(\ref{eq:kappa_t})), nonsingular solutions with accelerated expansion at late times are found for both positive and negative values of $\kappa_0$ and $\epsilon$. As an illustration, plots of the scale factor $a(t)$ and the corresponding $\kappa(t)$ for $\kappa_0>0$ and $\epsilon<0$, are shown in Figs.~\ref{fig:a_vac} and \ref{fig:k_vac}. From Figs.~\ref{fig:y_vac} and \ref{fig:doty_vac}, we note that $y\rightarrow 0$ and $\dot{y}\rightarrow 0$ at late times. During this phase, $\frac{\dot{a}}{a}\simeq \frac{\dot{\kappa}}{2\kappa}$ [from Eq.~(\ref{eq:kappavarn_cosmo})]. As a result, $a\propto \sqrt{\vert \kappa \vert}$, or $a\propto \exp(\mu t/2)$, since  $ \vert \epsilon \exp(\mu t) \vert \gg \vert\kappa_0\vert $ at large $t$ for $\mu>0$. For $\epsilon>0$, $y\rightarrow 2$ and $\dot{y}\rightarrow 0$, and therefore $a\propto \exp(\mu t/6)$ at large $t$. Thus the scale factor approaches de Sitter expansion stage at late times for $\mu>0$. As we will see later, for $\epsilon<0$, a similar reasoning applies to the Universe filled with dust or radiation, which also approaches the de Sitter expansion stage at late times when $\vert \kappa \rho\vert \sim 0 $. This becomes clear from the numerical plots shown later. Although we get an expression for asymptotic behavior of $a(t)$ at late times, we need to solve the system of equations numerically to obtain the full solution. 

\subsection{$p=0$, dust}
\noindent For dust ($p=0$), $\rho=\rho_0/a^3$, where $\rho_0$ is a constant. Thus, $U$ and $V$ become
\begin{equation}
U=\frac{(2-y)^{3/2}}{\sqrt{y+\kappa \rho}}~\quad \mbox{and}~\quad V=\sqrt{(y+\kappa\rho)(2-y)}.
\label{eq:UV_p=0}
\end{equation}   
We define two new functions
\begin{eqnarray}
F_1&:=&1+2U-3\frac{U}{V}\nonumber\\
&=&1+\frac{2(2-y)^{3/2}}{\sqrt{y+\kappa \rho}}-\frac{3(2-y)}{y+\kappa \rho},
\label{eq:F1}
\end{eqnarray}
and
\begin{eqnarray}
\beta &:=& \frac{\dot{a}}{a}+\frac{\dot{V}}{2V}\nonumber\\
&=&\frac{1}{4(y+\kappa \rho)} \left[\frac{\dot{y}(2-2y-\kappa \rho)}{2-y}+\frac{\dot{a}}{a}(4y+\kappa \rho)+\mu (\kappa-\kappa_0)\rho \right],
\label{eq:beta_p=0}
\end{eqnarray}
where we have used the Eq.~(\ref{eq:kappa_t}). Using Eqs.~(\ref{eq:kappavarn_cosmo}), (\ref{eq:F1}), and (\ref{eq:beta_p=0}), we get
\begin{eqnarray}
\frac{\dot{a}}{a}&=&\frac{(y+\kappa \rho)\left[4\beta (2-y)+\mu (\kappa -\kappa_0)\left\lbrace \frac{(2y+\kappa \rho -2)F_1}{2\kappa}-\rho\right\rbrace\right]}{4(2y^2-4y+3)+\kappa \rho (5y-4)}\nonumber\\
&\equiv & H(a,\kappa, y, \beta)\, ,
\label{eq:a_dot_p=0_ch6}
\end{eqnarray}
and
\begin{eqnarray}
\dot{y}&=&-6(y-1)H + \mu (\kappa -\kappa_0)\left[\frac{(y+\kappa \rho)F_1}{2\kappa}-\rho\right]\nonumber\\
&\equiv&F_y(a,\kappa, y, \beta) \, .
\label{eq:y_dot}
\end{eqnarray}
Furthermore, making use of Eq.~(\ref{eq:R_eq}) we get 
\begin{eqnarray}
\dot{\beta}&=& \frac{1}{12\beta}\frac{d}{dt}\left(\frac{F_1}{\kappa}\right)\nonumber\\
&\equiv & F_{\beta}(a,\kappa, y, \beta).
\label{eq:beta_dot}
\end{eqnarray}
Using Eq.~(\ref{eq:F1}), we compute
\begin{eqnarray}
F_{\beta}&=& \frac{1}{12\beta \kappa}\left[\frac{\partial F_1}{\partial y}F_y-3\rho \frac{\partial F_1}{\partial \rho}H + \mu (\kappa -\kappa_0) \left(\frac{\partial F_1}{\partial \kappa}-\frac{F_1}{\kappa} \right) \right],
\end{eqnarray}
where $H$ and $F_y$ are given by the rhs of the Eqs.~(\ref{eq:a_dot_p=0_ch6}) and (\ref{eq:y_dot}). We solve numerically the system of first order ordinary differential equations (ODEs) (\ref{eq:a_dot_p=0_ch6}), (\ref{eq:y_dot}), and (\ref{eq:beta_dot}) along with Eq.~(\ref{eq:kappa_t}). We need only three initial conditions $a(0)$, $y(0)$, and $\kappa(0)$. Then $\beta(0)$ is fixed, $\beta(0)=\pm \beta_0$, where $\beta_0^2=(F_1/{6\kappa})\vert_{\lbrace a(0),y(0),\kappa(0)\rbrace}$. However, in our solution, we choose an appropriate  sign for $\beta(0)$ such that $H(0)>0$. We also choose $\kappa(0)\sim \kappa_0$ and $y(0)\sim 1$ so that we start from an 
EiBI regime of the solution.

\subsubsection{$\mu>0$ case:}

\begin{figure}[!htbp]
\centering
\subfigure[]
{
\mbox{\epsfig{file=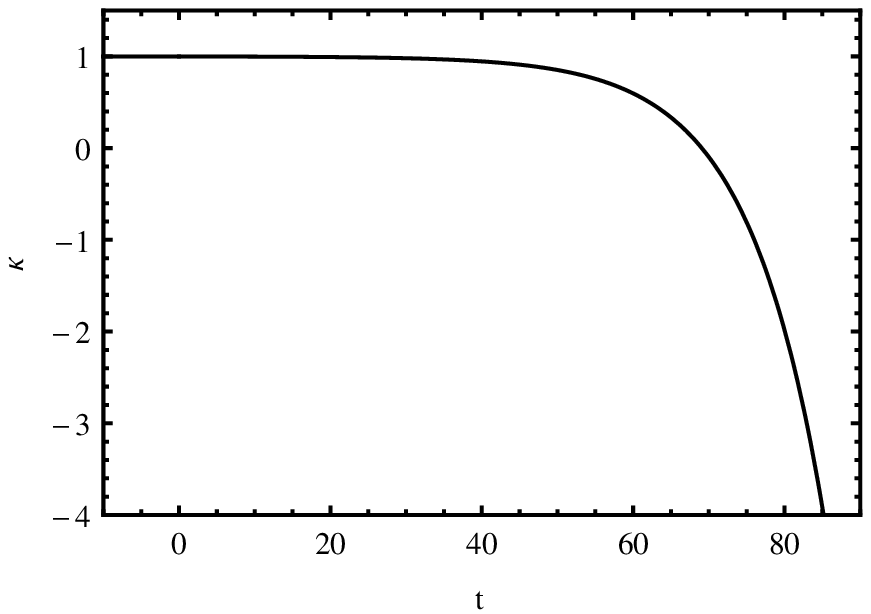,width=0.45\textwidth,angle=360}}
\label{fig:k_101}
}
\subfigure[]{
\mbox{\epsfig{file=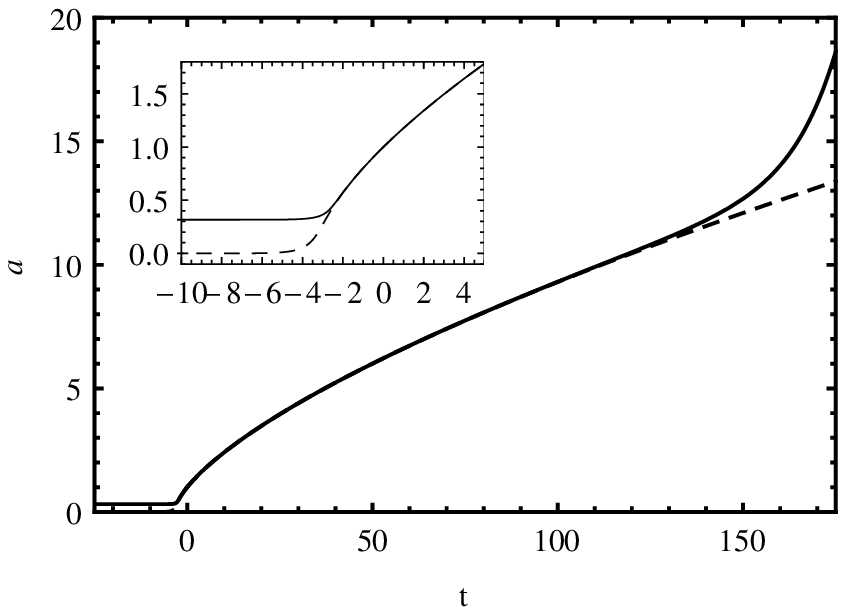,width=0.45\textwidth,angle=360}}
\label{fig:a_101}
}\\
\subfigure[]
{
\mbox{\epsfig{file=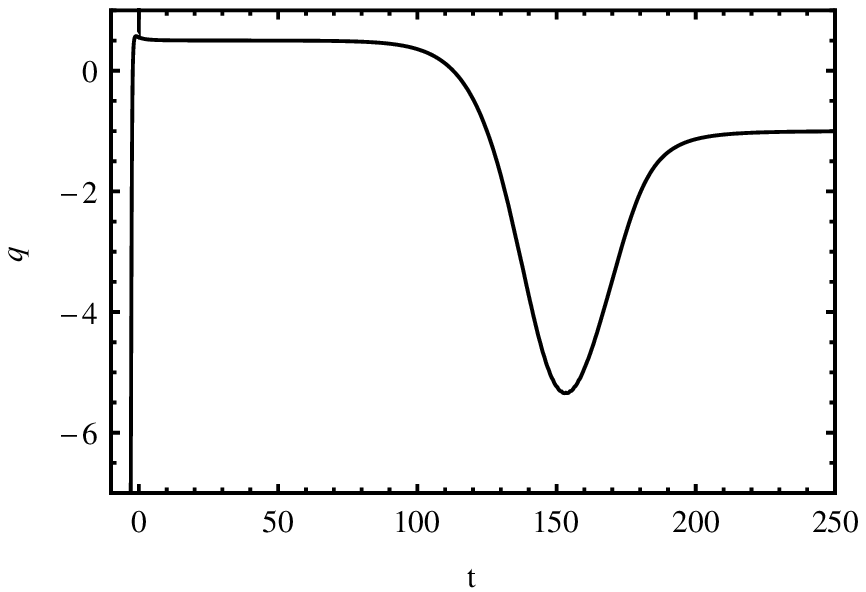,width=0.45\textwidth,angle=360}}
\label{fig:q_101}
}
\subfigure[]{
\mbox{\epsfig{file=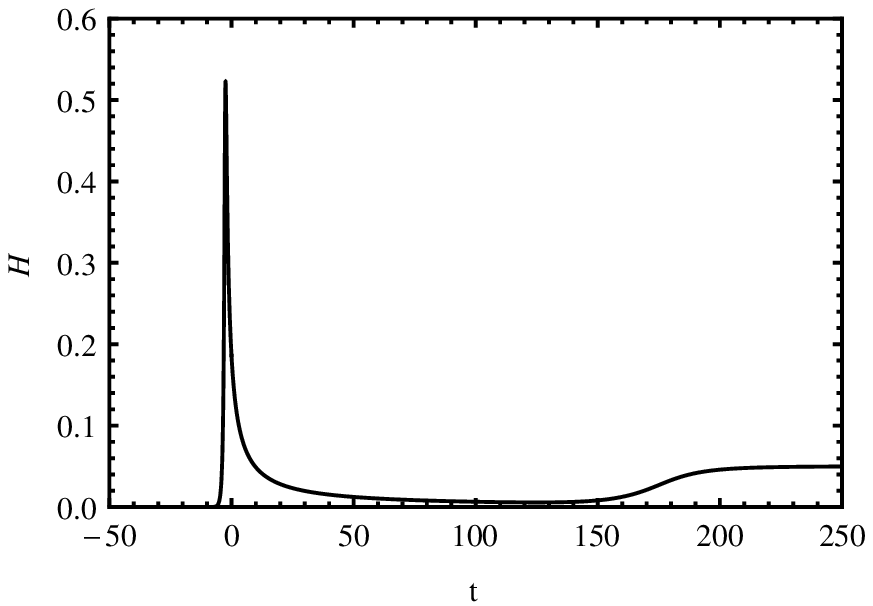,width=0.45\textwidth,angle=360}}
\label{fig:h_101}}\\
\subfigure[]
{
\mbox{\epsfig{file=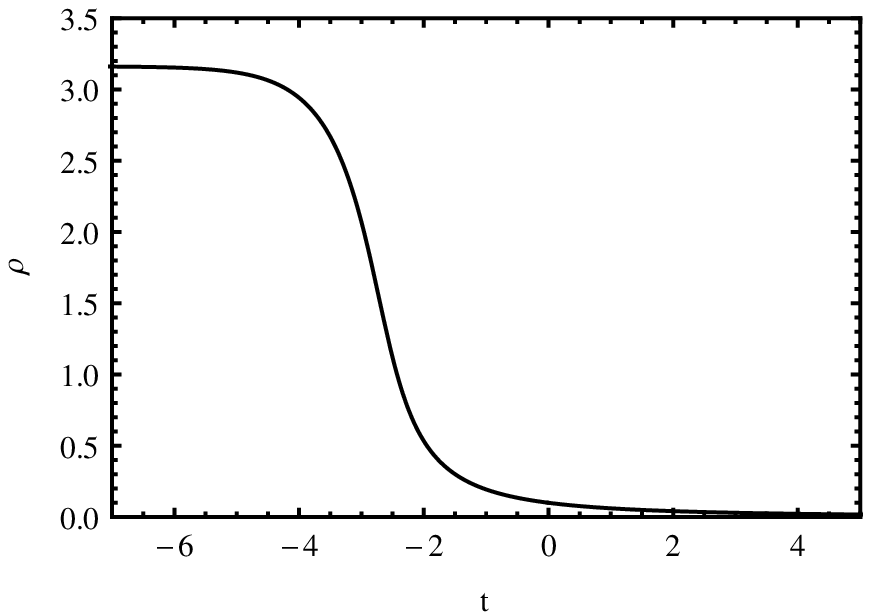,width=0.45\textwidth,angle=360}}
\label{fig:rho_101}
}
\subfigure[]{
\mbox{\epsfig{file=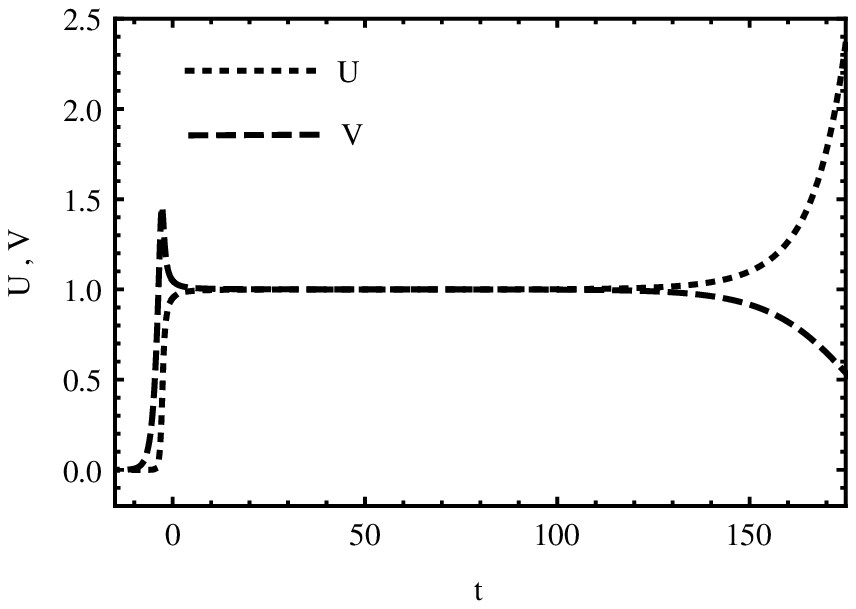,width=0.45\textwidth,angle=360}}
\label{fig:UV_101}}
\caption{Plot of (a) $\kappa(t)$, (b) $a(t)$, (c) $q(t)$, (d) $H(t)$, (e) $\rho(t)$, and (f) $U(t)$ (dotted line), $V(t)$ (dashed line) for $\kappa_0 >0$, $\epsilon<0$, and $\mu>0$. The parameters used are $\kappa_0=1$, $\mu=0.1$, and $\rho_0=0.1$. We choose $a(0)=1$, $y(0)=1.001$, $\kappa(0)=0.999$ for the numerical solution. The dashed black curve in (b) corresponds to the EiBI solution with $\kappa=\kappa_0$ (constant). Equation of state (EOS), $p=0$. }
\label{fig:aq_101}
\end{figure}

\noindent For $\mu>0$, we may choose $\kappa_0$ and $\epsilon$ as either positive or 
negative. From the analysis of our numerical solutions, we found that the 
solutions are nonsingular only for $\epsilon <0$. For $\kappa_0>0$ and 
$\epsilon <0$, $\kappa$ decreases with the increase in time, changes sign 
from positive to negative, and becomes more and more negative with time 
(see Fig.~\ref{fig:k_101}). In this case, the early Universe undergoes a 
loitering phase (see Fig.~\ref{fig:a_101}). This is similar to the case of a 
constant positive $\kappa$ in EiBI gravity \citep{banados,cho}. However, we 
note that the scale factor $a(t)$ never goes to zero unlike the case in EiBI 
gravity, where $a\rightarrow 0$ as $t\rightarrow -\infty$ for a dust-filled 
Universe \citep{cho}. This is demonstrated in the inset of Fig.~\ref{fig:a_101}, where the dashed curve denotes the $\kappa=\kappa_0$ case and the solid curve denotes the $\kappa(t)$ case. The plot also demonstrates the accelerated 
expansion of the Universe at late times. This feature is absent in EiBI theory 
and GR, where we see deceleration of the Universe at late times for $p=0$. 
Figure~\ref{fig:q_101} shows the plot of the deceleration parameter $q$. We know that, in GR, for a matter (dust-) dominated Universe, $a(t)\propto t^{2/3}$ and, consequently, $q=0.5$. In the plot of $q$ (Fig.~\ref{fig:q_101}), we see that there are large variations from the value in GR, both at early and late times. 
In the intermediate range of time scale ($t\sim 0-100$), we see a GR-like phase. We also note that the Universe asymptotically approaches a de Sitter expansion phase ($q\rightarrow -1$) at late times (for $t>200$ in the plot). This fact 
is also evident from Fig.~\ref{fig:h_101}, where the Hubble function $H$ 
becomes almost a constant at late times. Figure~\ref{fig:rho_101} shows that 
there is a finite maximum energy density or, conversely, a nonzero minimum 
scale factor. This is unlike the case in EiBI gravity, where 
$\rho\rightarrow \infty$ as $t\rightarrow -\infty$ for the $p=0$ case  
\citep{cho}. From Fig.~\ref{fig:UV_101}, we note that $U\sim 1$ and $V\sim 1$ during the GR-like phase (i.e. $t\sim 0-100$) but varies largely at both the 
early ($t<0$) and late times ($t>100$).

\begin{figure}[htbp!]
\centering
\subfigure[]{
\mbox{\epsfig{file=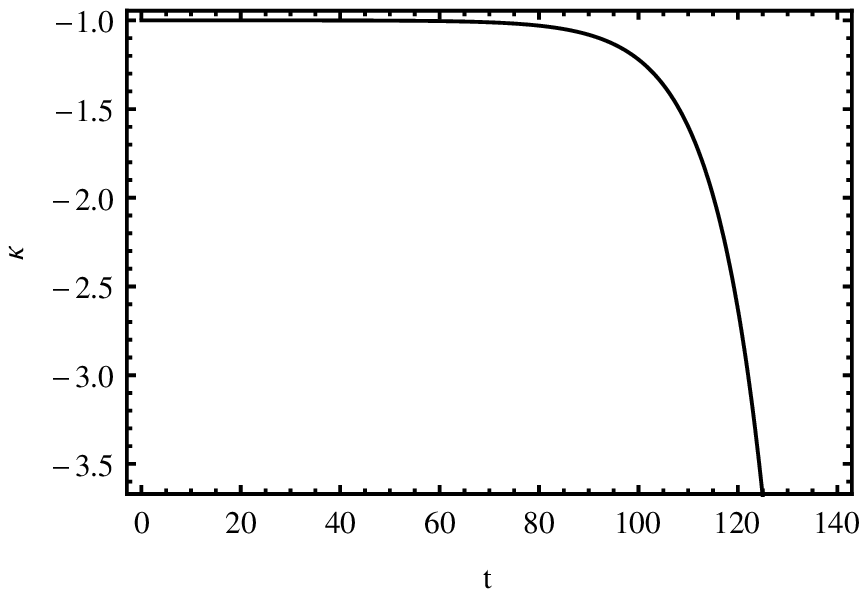,width=0.45\textwidth,angle=360}}
\label{fig:k_001}}
\subfigure[]{
\mbox{\epsfig{file=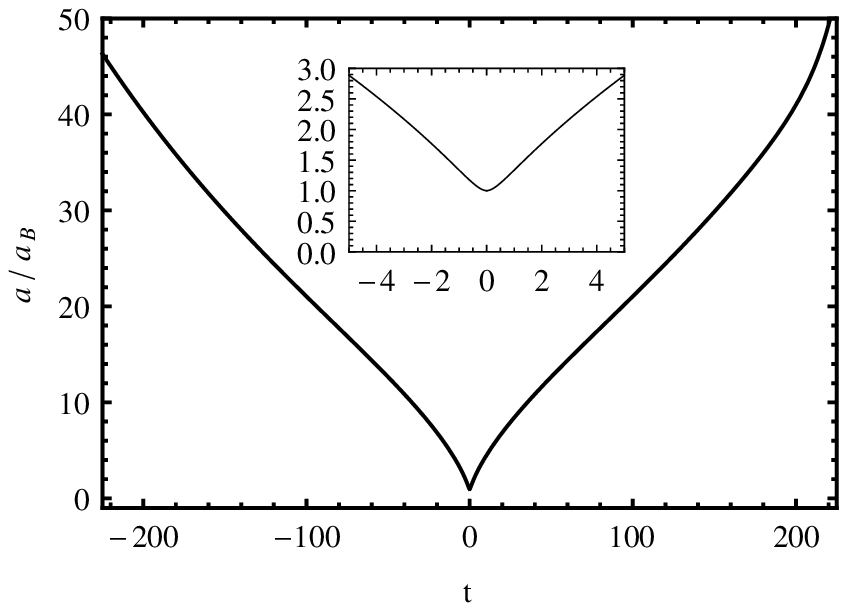,width=0.45\textwidth,angle=360}}
\label{fig:a_001}}\\
\subfigure[]
{
\mbox{\epsfig{file=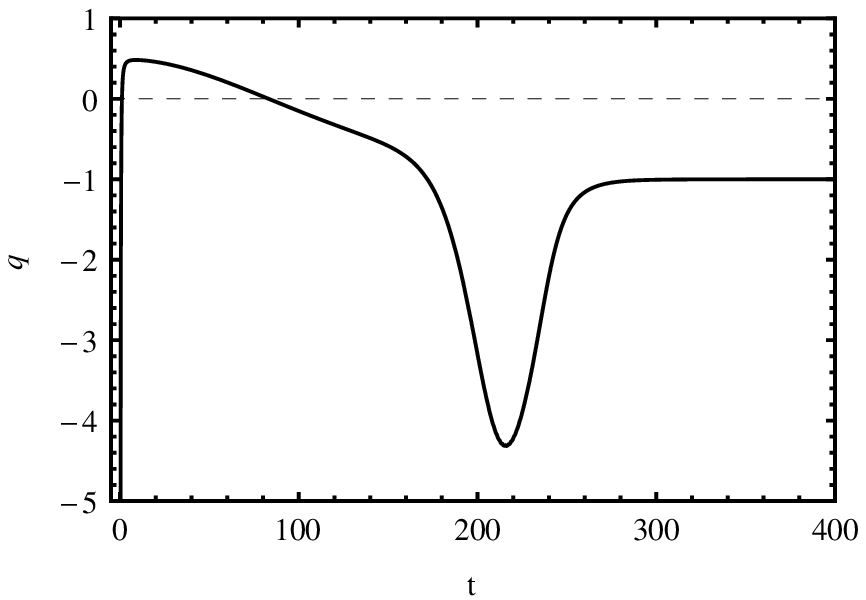,width=0.45\textwidth,angle=360}}
\label{fig:q_001}
}
\subfigure[]{
\mbox{\epsfig{file=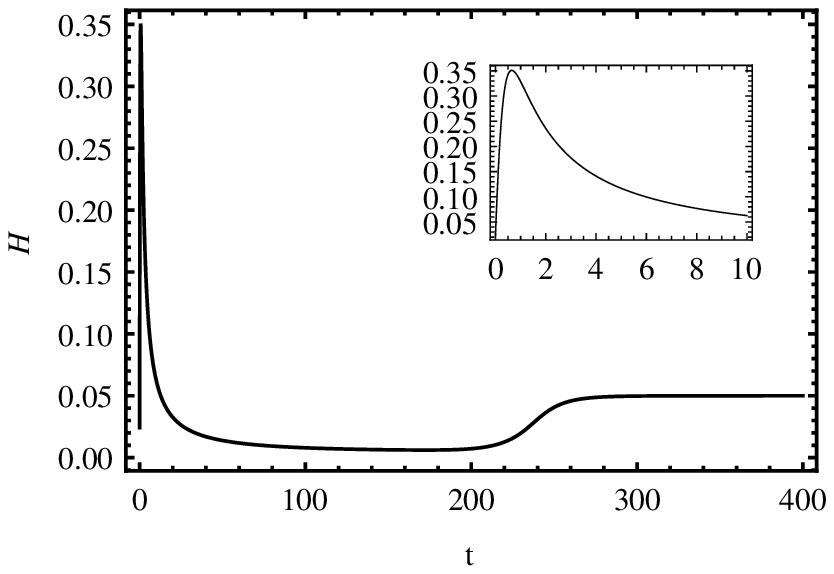,width=0.45\textwidth,angle=360}}
\label{fig:h_001}}\\
\subfigure[]
{
\mbox{\epsfig{file=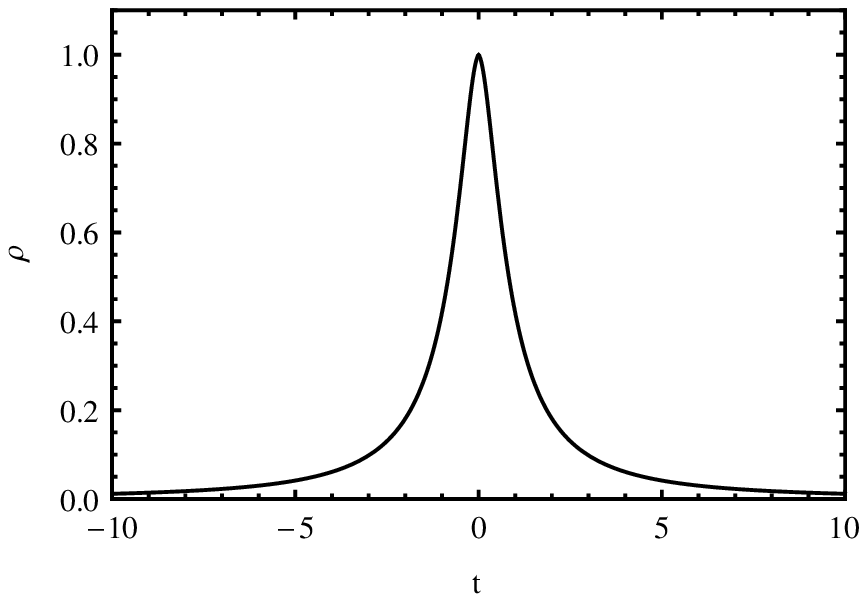,width=0.45\textwidth,angle=360}}
\label{fig:rho_001}
}
\subfigure[]{
\mbox{\epsfig{file=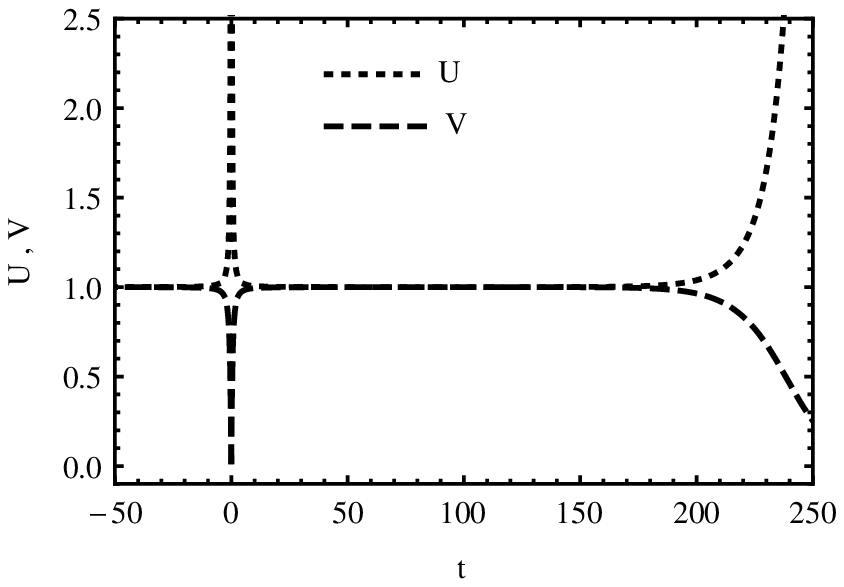,width=0.45\textwidth,angle=360}}
\label{fig:UV_001}}
\caption{Plot of (a) $\kappa(t)$, (b) $a(t)$, (c) $q(t)$, (d) $H(t)$, (e) $\rho(t)$, and (f) $U(t)$ and $V(t)$ for $\kappa_0 <0$, $\epsilon<0$, and $\mu>0$. The parameters used are $\kappa_0=-1$, $\mu=0.1$, and $\rho_0=0.01$. We choose $a(0)=(-\kappa_0\rho_0)^{1/3}$, $y(0)=0.9999$, $\kappa(0)=-1.00001$ for the numerical solution. EOS, $p=0$. Evolution of $a(t)$ and $H(t)$ near the {\em bounce} are shown in the insets of (b) and (d). $q$ and $U$ diverge at bounce. }
\label{fig:aq_001}
\end{figure}

\noindent For $\kappa_0<0$ and $\epsilon<0$, $\kappa$ is always negative and, with increase in time, $\vert \kappa \vert$ increases (see Fig.~\ref{fig:k_001}). In this 
case, the Universe undergoes a bounce instead of a loitering phase at early 
times. This is similar to EiBI gravity. Late-time accelerated expansion of 
the Universe  occurs after a deceleration which immediately follows the 
bounce. This feature is understood through the plots of the scale factor $a(t)$ in Fig.~\ref{fig:a_001} and the deceleration parameter $q$ in Fig.~\ref{fig:q_001}. Here also, the Universe reaches, asymptotically, a de Sitter expansion at late times ($t>280$ in the plots for $q$ and $H$).   

\noindent The case $\epsilon>0$ is not shown here through plots.
We have checked that our numerical solutions  reveal an early loitering phase for $\kappa_0>0$ and a bounce for $\kappa_0<0$, as expected
($\kappa$ approaches the constant value $\kappa_0$ at early times, i.e. 
$\kappa\rightarrow \kappa_0$ as $t\rightarrow -\infty$). 
Thus, the early Universe is still nonsingular. However, in both the cases, 
for $\epsilon>0$, a singularity appears at a finite future time $t_f$ where 
$H$diverges ($H\rightarrow -\infty$ as $t\rightarrow t_f$). The scale factor 
$a(t)$ and the energy density $\rho(t)$ though remain finite at $t_f$. This is a type-III (Big Freeze) singularity according to the classification given in \cite{nojiri2005,nojiri2005b} and it yields a geodesically complete spacetime that 
does not necessarily crush/destroy physical observers \cite{jimenez2016}. 

\subsubsection{$\mu<0$ case:}

\begin{figure}[htbp!]
\centering
\subfigure[]{
\mbox{\epsfig{file=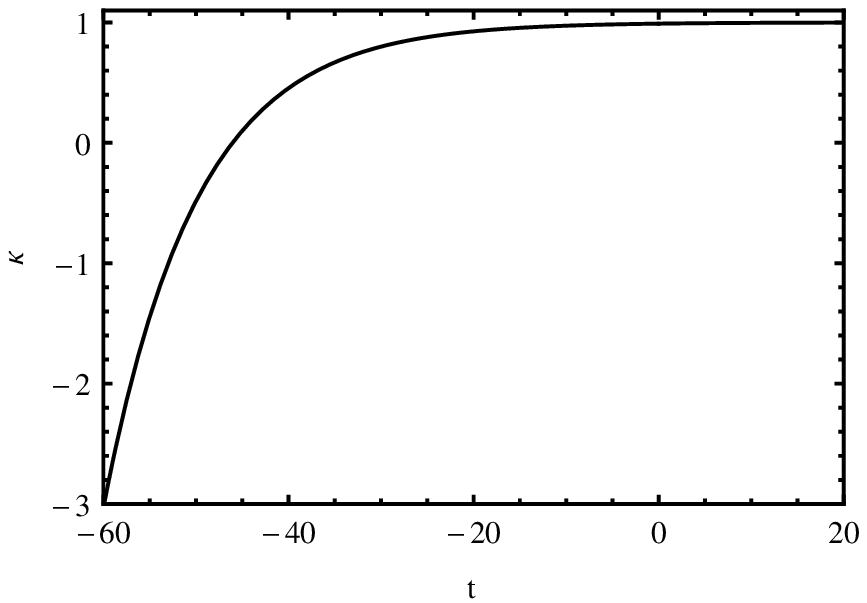,width=0.45\textwidth,angle=360}}
\label{fig:k_100}}
\subfigure[]{
\mbox{\epsfig{file=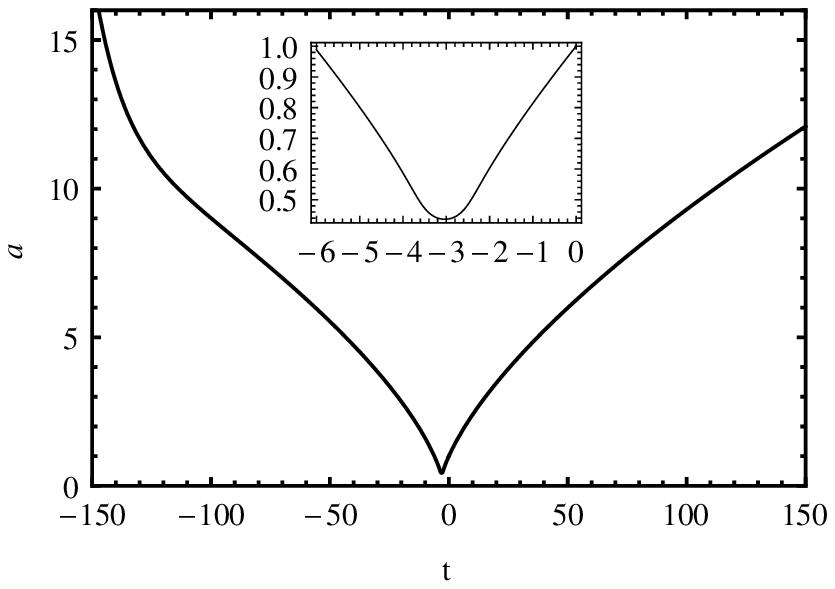,width=0.45\textwidth,angle=360}}
\label{fig:a_100}}\\
\subfigure[]
{
\mbox{\epsfig{file=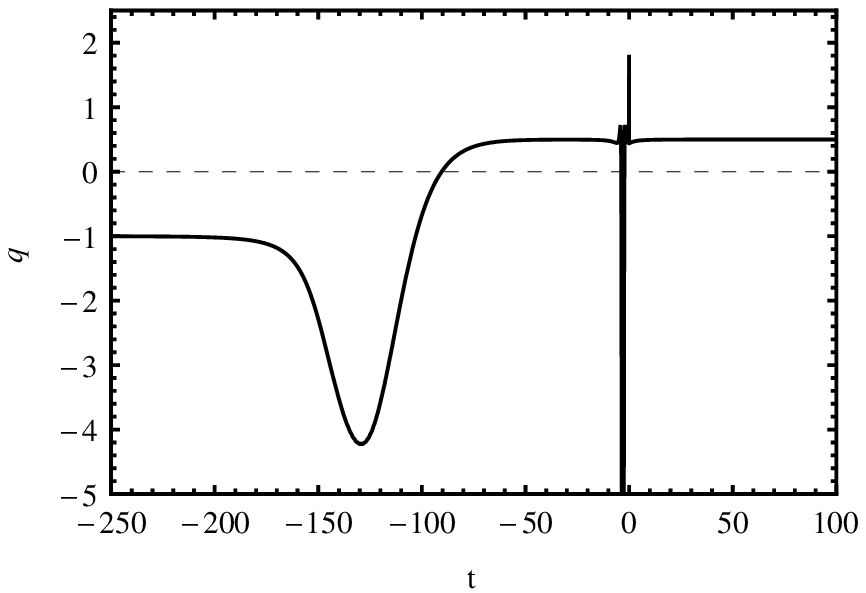,width=0.45\textwidth,angle=360}}
\label{fig:q_100}
}
\subfigure[]{
\mbox{\epsfig{file=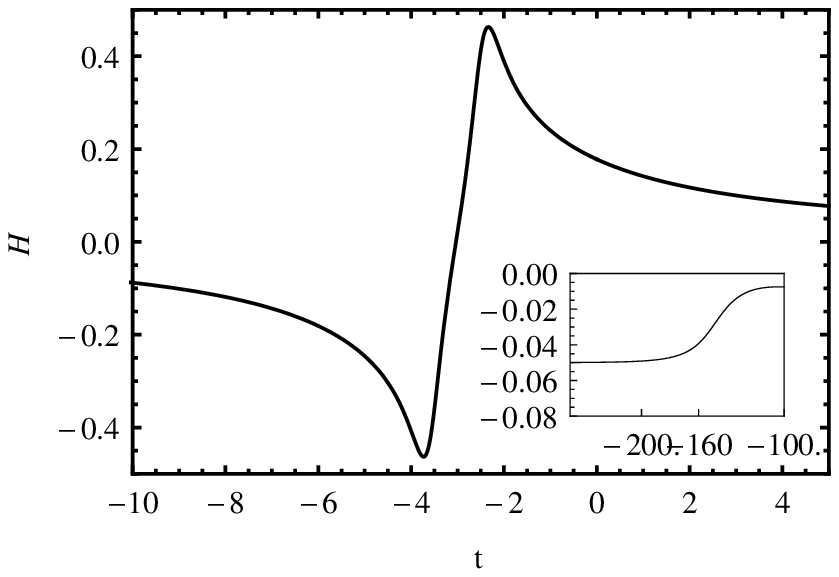,width=0.45\textwidth,angle=360}}
\label{fig:h_100}}\\
\subfigure[]
{
\mbox{\epsfig{file=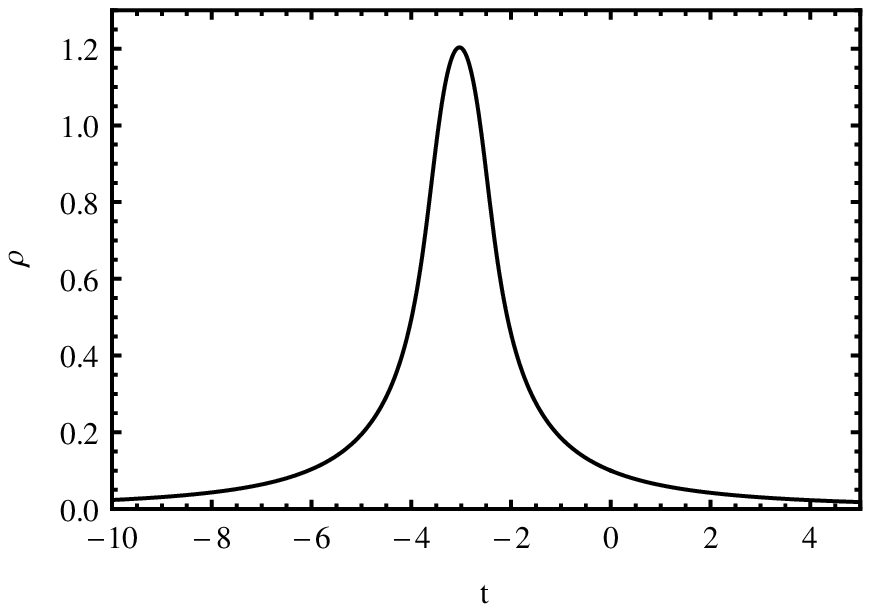,width=0.45\textwidth,angle=360}}
\label{fig:rho_100}
}
\subfigure[]{
\mbox{\epsfig{file=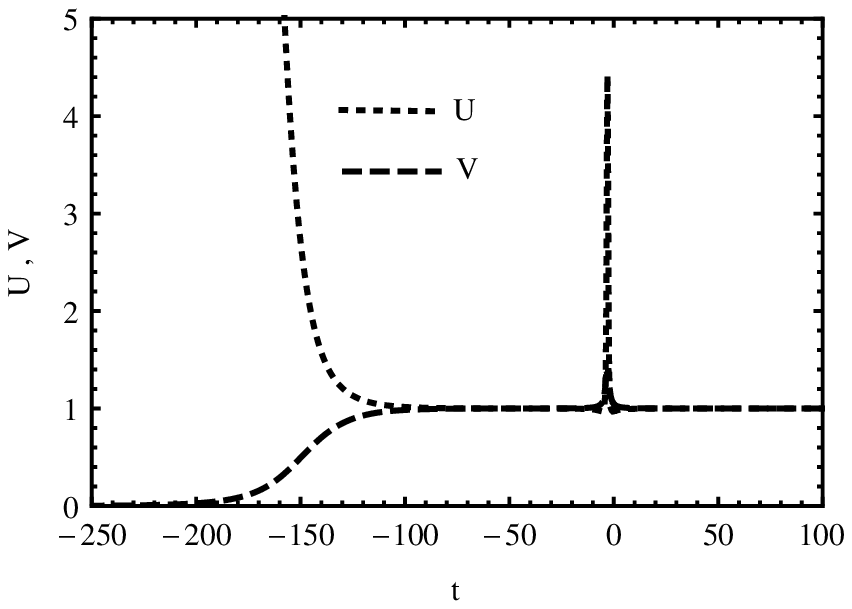,width=0.45\textwidth,angle=360}}
\label{fig:UV_100}}
\caption{Plot of (a) $\kappa(t)$, (b) $a(t)$, (c) $q(t)$, (d) $H(t)$, (e) $\rho(t)$, and (f) $U(t)$ and $V(t)$ for $\kappa_0 >0$, $\epsilon<0$, and $\mu<0$. The parameters used are $\kappa_0=1$, $\mu=-0.1$, and $\rho_0=0.1$. We choose $a(0)=1$, $y(0)=0.99$, $\kappa(0)=0.99$ for the numerical solution. EOS, $p=0$. Evolution of $a(t)$ about the {\em bounce} is shown clearly in inset of (b). Inset of (d) shows that $H(t)$ approaches a constant negative value as $t\rightarrow-\infty$. $q$ diverges to $-\infty$ at bounce.}
\label{fig:aq_100}
\end{figure}

\begin{figure}[htbp!]
\centering
\subfigure[]{
\mbox{\epsfig{file=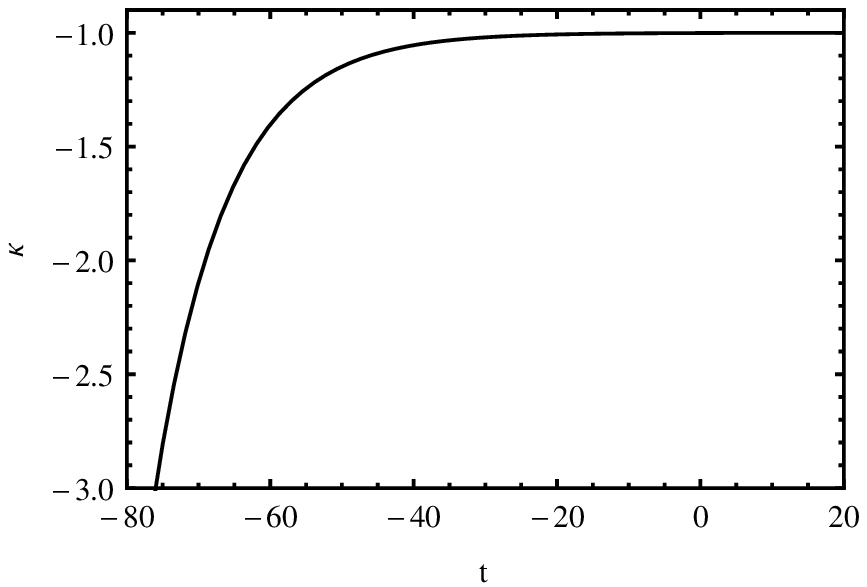,width=0.45\textwidth,angle=360}}
\label{fig:k_000}}
\subfigure[]{
\mbox{\epsfig{file=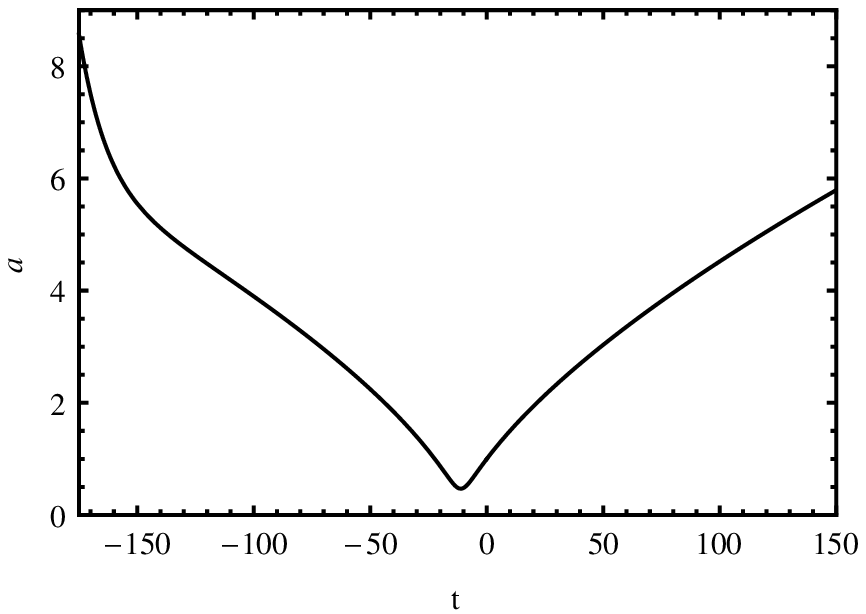,width=0.45\textwidth,angle=360}}
\label{fig:a_000}}\\
\subfigure[]
{
\mbox{\epsfig{file=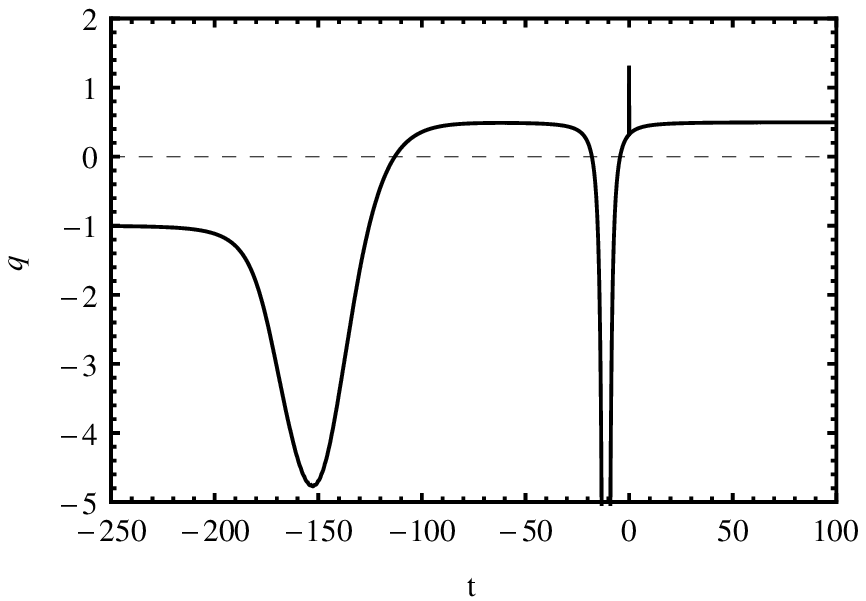,width=0.45\textwidth,angle=360}}
\label{fig:q_000}
}
\subfigure[]{
\mbox{\epsfig{file=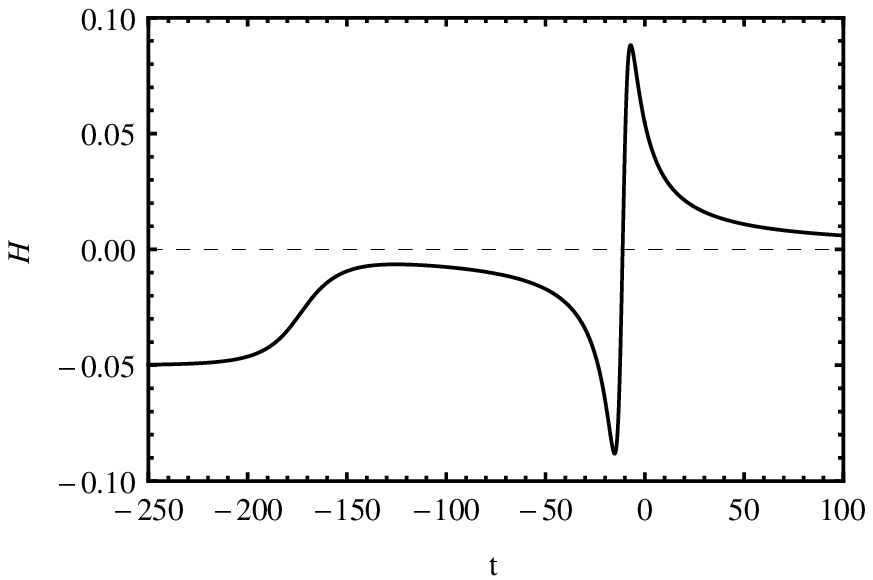,width=0.45\textwidth,angle=360}}
\label{fig:h_000}}\\
\subfigure[]
{
\mbox{\epsfig{file=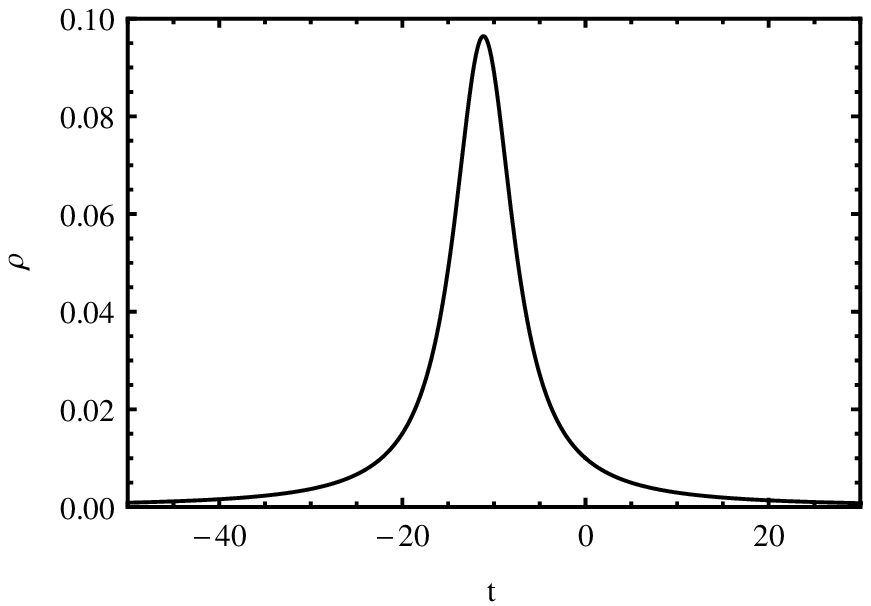,width=0.45\textwidth,angle=360}}
\label{fig:rho_000}
}
\subfigure[]{
\mbox{\epsfig{file=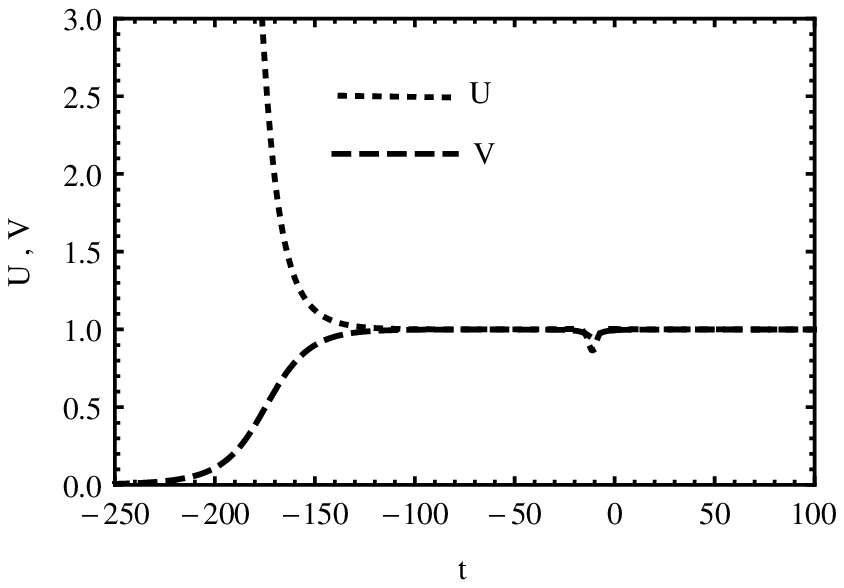,width=0.45\textwidth,angle=360}}
\label{fig:UV_000}}
\caption{Plot of (a) $\kappa(t)$, (b) $a(t)$, (c) $q(t)$, (d) $H(t)$, (e) $\rho(t)$, and (f) $U(t)$ and $V(t)$ for $\kappa_0 <0$, $\epsilon<0$, and $\mu<0$. The parameters used are $\kappa_0=-1$, $\mu=-0.1$, and $\rho_0=0.01$. We choose $a(0)=1$, $y(0)=1.001$, $\kappa(0)=-1.001$ for the numerical solution. EOS $p=0$. $q$ diverges to $-\infty$ at bounce.}
\label{fig:aq_000}
\end{figure}

\noindent For $\mu<0$, $\kappa$ approaches $\kappa_0$ asymptotically as 
$t\rightarrow \infty$. Thus, the solutions tend to the EiBI solutions for 
constant $\kappa_0$, at large $t$. In this case also, a nonsingular Universe 
is found for $\epsilon<0$. However, we do not see a loitering early stage 
for $\kappa_0>0$. A bounce occurs for both $\kappa_0>0$ and $\kappa_0<0$. 
We note that an accelerated contraction precedes the decelerated contraction, 
before the bounce occurs. These features are shown in Figs.~\ref{fig:aq_100} and \ref{fig:aq_000}.  Figs.~\ref{fig:q_100} and \ref{fig:q_000} show that $q\rightarrow -1$ as $t\rightarrow -\infty$. $H$ approaches a constant negative value during this period (see the inset of Fig.~\ref{fig:h_100} and the Fig.~\ref{fig:h_000}). Also, we see that $q\sim 0.5$ in between the bounce and accelerated contraction phase, and throughout the future time after the bounce. Thus, evolution of the scale factor is GR like during these periods. It may also be noted 
that $U\sim 1$ and $V\sim 1$ in these phases.  

\noindent
The solutions are singular for $\epsilon>0$. Therefore, we only mention the results, but do not show the plots. For $\kappa_0>0$ and $\epsilon>0$, there may exist a big bang singularity. The Universe may also begin with a singularity at $t=-t_f$ where $H$ diverges ($H(-t_f)\rightarrow \infty$), but $a$ and $\rho$ are finite. The last kind of singularity always occurs for $\kappa_0<0$ and $\epsilon>0$. This is similar to the type-III singularity mentioned earlier \cite{nojiri2005,nojiri2005b}. 
However, future singularities do not occur unlike the case $\mu>0$ and $\epsilon>0$.   

\subsection{$p=\rho/3$ case}
We now turn to a Universe filled with the radiation ($p=\rho/3$). We have 
$\rho=\rho_0/a^4$, and
\begin{equation}
U=\frac{(2-y-\kappa \rho/3)^{3/2}}{\sqrt{y+\kappa \rho}}~\quad \mbox{and}~\quad V=\sqrt{(y+\kappa \rho)(2-y-\kappa \rho/3)}.
\label{eq:UV_rad_ch6}
\end{equation}
Thus, $F_1$, $H$, and $F_{\beta}$ are now given as,
\begin{eqnarray}
F_1&=&\frac{4y-6+2\kappa \rho+ 2\sqrt{(y+\kappa \rho)(2-y-\kappa \rho/3)^3}}{y+\kappa \rho}, \label{eq:F1_rad}\\
H&=& \frac{(y+\kappa \rho)\left[ 4\beta\left(2-y-\frac{\kappa \rho}{3}\right)+\mu(\kappa -\kappa_0)\left\lbrace \frac{F_1}{\kappa}\left(y+\frac{2\kappa \rho}{3}-1\right)-\frac{2\rho}{3} \right \rbrace\right]}{4\left[\left(2y^2 -4y +3\right) + 2\kappa \rho \left(y-1\right) + \frac{\kappa^2\rho^2}{3}\right]},
\label{eq:H_rad}\\
F_{\beta}&=&\frac{1}{12\beta \kappa}\left[\frac{\partial F_1}{\partial y}F_y-4\rho \frac{\partial F_1}{\partial \rho}H + \mu (\kappa -\kappa_0) \left(\frac{\partial F_1}{\partial \kappa}-\frac{F_1}{\kappa} \right) \right].
\label{eq:F_beta_rad}
\end{eqnarray}
The expression of $F_y$ remains unchanged (\ref{eq:y_dot}). We solve the 
system of ODEs, $\dot{a}=aH$, $\dot{y}=F_y$, $\dot{\beta}=F_{\beta}$, 
and $\dot{\kappa}=\mu(\kappa -\kappa_0)$ numerically.
 
\begin{figure}[htbp!]
\centering
\subfigure[]{
\mbox{\epsfig{file=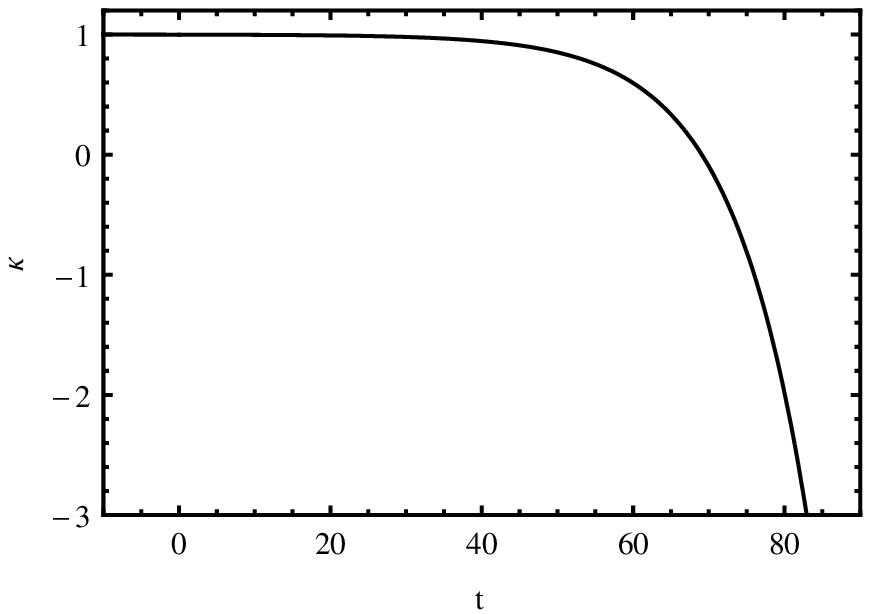,width=0.45\textwidth,angle=360}}
\label{fig:krad_101}}
\subfigure[]{
\mbox{\epsfig{file=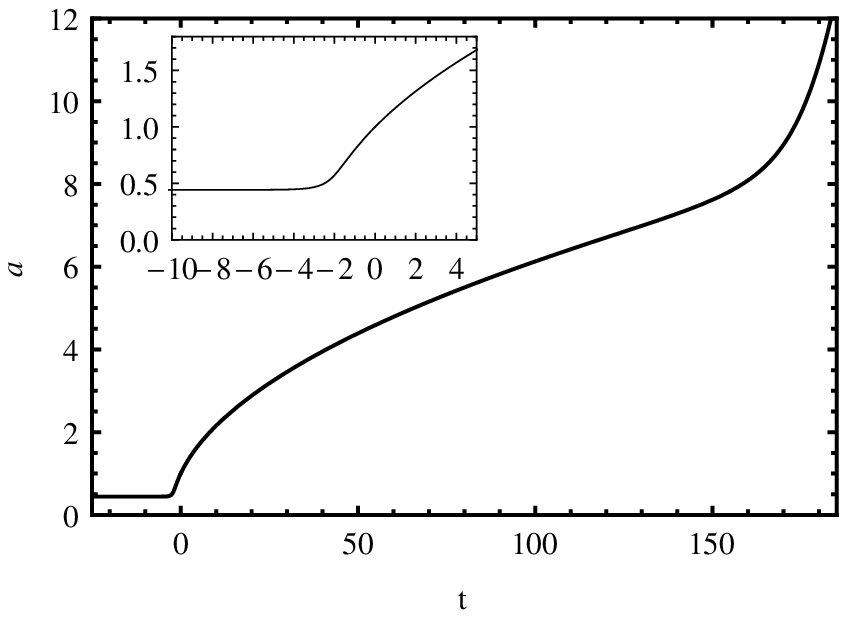,width=0.45\textwidth,angle=360}}
\label{fig:arad_101}}\\
%\subfigure[]
%{
%\mbox{\epsfig{file=Figure/qrad_101.eps,width=0.45\textwidth,angle=360}}
%\label{fig:qrad_101}
%}
%\subfigure[]{
%\mbox{\epsfig{file=Figure/hrad_101.eps,width=0.45\textwidth,angle=360}}
%\label{fig:hrad_101}}\\
%\subfigure[]
%{
%\mbox{\epsfig{file=Figure/rhorad_101.eps,width=0.45\textwidth,angle=360}}
%\label{fig:rhorad_101}
%}
%\subfigure[]{
%\mbox{\epsfig{file=Figure/UVrad_101.eps,width=0.45\textwidth,angle=360}}
%\label{fig:UVrad_101}}
\caption{Plot of  (a) $\kappa(t)$ and (b) $a(t)$  for $\kappa_0 >0$, $\epsilon<0$, and $\mu>0$. The parameters used are $\kappa_0=1$, $\mu=0.1$, and $\rho_0=0.1$. We choose $a(0)=1$, $y(0)=1.001$, $\kappa(0)=0.999$ for the numerical solution. EOS, $p=\rho/3$. Inset of (b) shows the {\em loitering} phase where $a(t)$ approaches a nonzero minimum value asymptotically at past.}
\label{fig:aqrad_101}
\end{figure}

\begin{figure}[htbp!]
\centering
\subfigure[]{
\mbox{\epsfig{file=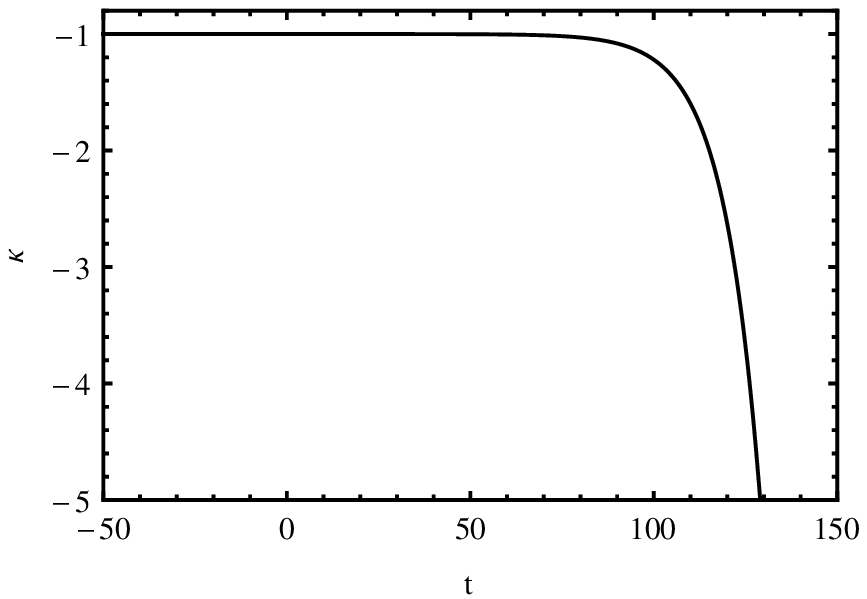,width=0.45\textwidth,angle=360}}
\label{fig:krad_001}}
\subfigure[]{
\mbox{\epsfig{file=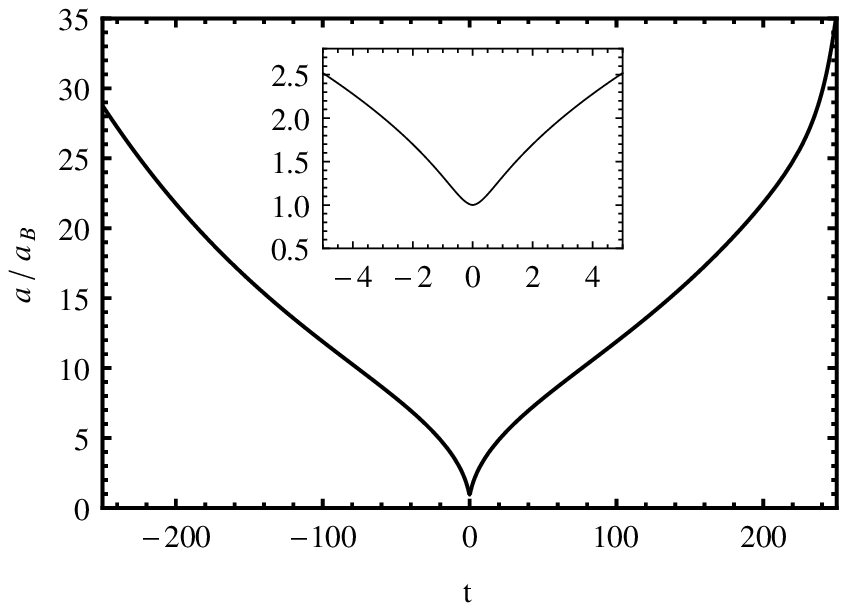,width=0.45\textwidth,angle=360}}
\label{fig:arad_001}}\\
%\subfigure[]
%{
%\mbox{\epsfig{file=Figure/qrad_001.eps,width=0.45\textwidth,angle=360}}
%\label{fig:qrad_001}
%}
%\subfigure[]{
%\mbox{\epsfig{file=Figure/hrad_001.eps,width=0.45\textwidth,angle=360}}
%\label{fig:hrad_001}}\\
%\subfigure[]
%{
%\mbox{\epsfig{file=Figure/rhorad_001.eps,width=0.45\textwidth,angle=360}}
%\label{fig:rhorad_001}
%}
%\subfigure[]{
%\mbox{\epsfig{file=Figure/UVrad_001.eps,width=0.45\textwidth,angle=360}}
%\label{fig:UVrad_001}}
\caption{Plot of (a) $\kappa(t)$ and (b) $a(t)$ for $\kappa_0 <0$, $\epsilon<0$, and $\mu>0$. The parameters used are $\kappa_0=-1$, $\mu=0.1$, and $\rho_0=0.01$. We choose $a(0)=(-\kappa_0\rho_0)^{1/4}$, $y(0)=0.9999$, $\kappa(0)=-1.00001$ for the numerical solution. Evolution of $a(t)$ near the {\em bounce} is 
shown in the inset of (b). EOS, $p=\rho/3$.}
\label{fig:aqrad_001}
\end{figure}

\noindent In the solutions, we note, qualitatively, the same features as seen 
in the $p=0$ case. A notable difference is that during the GR-like phases, 
$q\sim 1$. This is due to the fact that, in GR, for a radiation filled Universe, $a(t)\propto t^{1/2}$. Here also, the nonsingular solutions are found for 
$\epsilon<0$, irrespective of the sign of $\mu$ and $\kappa_0$.  
We illustrate some of the nonsingular 
scale factors through the plots in Figs.~\ref{fig:aqrad_101} 
and \ref{fig:aqrad_001}. %, \ref{fig:aqrad_100}, and \ref{fig:aqrad_000}. 
 
\noindent Apart from dust and radiation, we have also looked at the
vacuum case. It turns out that for $\kappa_0>0, \epsilon<0,\mu>0$,
the solution for the scale factor is qualitatively the same as in the
$p=0$ or $p=\frac{\rho}{3}$ cases. However, with $\kappa_0<0, \epsilon<0,\mu>0$
we do not obtain a bounce but a big-bang singularity.

\section{Conclusions}
In this article, we have explored the possibility of $\kappa$, the Born-Infeld parameter in EiBI gravity, being a real scalar field. 
In this way, we have proposed  a new theory of gravity by extending 
EiBI gravity in a manner similar to scalar-tensor theories. 
The action, equations of motion, energy-momentum conservation 
and the Newtonian limit of the theory have all been worked out.  
 
\noindent In order to derive some of the consequences of this new theory, 
we studied cosmology as an example. After choosing a specific form of $\kappa(t)$, we solved the field equations numerically 
for spatially flat FRW spacetimes with (i) dust ($p=0$) and 
(ii) radiation ($p=\rho/3$) as matter. In the case of the original 
EiBI theory (i.e. with a constant $\kappa$), we know that the solutions 
lead to a nonsingular early Universe, with a loitering phase for $\kappa>0$ and a bounce for $\kappa<0$. Further, the solutions reduce to those in GR at late 
times. In our work here, the choice of the scalar $\kappa(t)=\kappa_0+\epsilon \exp(\mu t)$ ( $\kappa_0$, $\epsilon$, and $\mu$ are constants) broadly leads to
qualitatively similar features for both $p=0$ and $p=\rho/3$. However
there are important additional features which arise. We summarize them 
pointwise below:
\begin{itemize}
\item Unlike the EiBI solutions, here, the solutions are not always 
nonsingular. For $\epsilon<0$, the solutions are nonsingular irrespective of 
the signs of $\kappa_0$ and $\mu$. The solutions with an early loitering phase of the Universe were found for $\kappa_0>0$, $\epsilon<0$, and $\mu>0$. All other nonsingular solutions have a bounce in the early Universe.
\item In EIBI gravity, with $p=0$, the early Universe is de Sitter when 
the constant $\kappa>0$. Therefore, $a\rightarrow 0$ at $t\rightarrow -\infty$. Consequently, $\rho\rightarrow \infty$ at $t\rightarrow -\infty$. 
In contrast, in our new theory, $a$ never goes to zero for the solution
with a loitering phase, and energy density remains finite for all $t$.
\item Late-time accelerated expansion of the Universe is an outcome 
for $\mu>0$ and $\epsilon<0$. 
The Universe becomes de Sitter ($q=-1$) asymptotically at large $t$.
Note that this happens without any additional matter but only via
the nature of $\kappa(t)$ and the structure of the theory.
\item In the vicinity of the minimum value of the scale factor, or 
conversely at high energy densities, there is a deviation in the time 
evolution of the scale factor from that in GR. There are deviations at 
large values of the scale factor or conversely, low energy densities, where 
we have noted acceleration. For intermediate values of the energy density, 
(or time scales), there exist GR-like phases, as expected.
 \end{itemize}     

\noindent Our work here is a glimpse of the interesting possibilities 
which may arise in this new theory. Much more work is surely required to
probe its feasibility. For example, we would  like to investigate whether 
there exists any nontrivial vacuum (or nonvacuum) spherically symmetric, 
static spacetimes in this new theory. This would be a major difference with
EiBI gravity where the vacuum solution is the Schwarzschild solution of GR. 
A different vacuum solution will affect the Solar System tests and put
bounds on the new parameters that are used in choosing $\kappa (t)$. 
Cosmological perturbation theory as well as the study of gravitational
waves in this theory might also be useful avenues to pursue 
in the context of this modified theory of gravity which encodes both
a Born-Infeld structure and a Brans-Dicke character in its formulation. For example, authors of \cite{escamilla} studied tensor perturbations about the homogeneous and isotropic cosmological background spacetimes of both bouncing and loitering nature, in EiBI theory. They found instabilities in the 
overall evolution, even though the background evolution is nonsingular and 
more so for the case of bouncing solutions as the background spacetimes. 
Whether such instabilities arise in this new theory too 
is an interesting question which requires detailed study.    
 
\noindent An important issue which must be dealt with is the origin of
the real scalar field $\kappa$. It is not appropriate to leave it as
an {\em ad hoc} entity. However, it is possible to speculate 
that such a scalar may have a higher-dimensional origin following 
work in the context of string theory and in braneworld models.

\bibliographystyle{elsarticle-num} 
\bibliography{reference}

\begin{thebibliography}{10}
\expandafter\ifx\csname url\endcsname\relax
  \def\url#1{\texttt{#1}}\fi
\expandafter\ifx\csname urlprefix\endcsname\relax\def\urlprefix{URL }\fi
\expandafter\ifx\csname href\endcsname\relax
  \def\href#1#2{#2} \def\path#1{#1}\fi

\bibitem{abbott}
B.~P. {Abbott}, R.~{Abbott}, T.~D. {Abbott}, M.~R. {Abernathy}, F.~{Acernese},
  K.~{Ackley}, C.~{Adams}, T.~{Adams}, P.~{Addesso}, R.~X. {Adhikari}, et~al.,
  {Tests of General Relativity with GW150914}, Physical Review Letters 116~(22)
  (2016) 221101.
\newblock \href {http://arxiv.org/abs/1602.03841} {\path{arXiv:1602.03841}},
  \href {http://dx.doi.org/10.1103/PhysRevLett.116.221101}
  {\path{doi:10.1103/PhysRevLett.116.221101}}.

\bibitem{Will2014}
C.~M. Will, The confrontation between general relativity and experiment, Living
  Reviews in Relativity 17~(1) (2014) 4.
\newblock \href {http://dx.doi.org/10.12942/lrr-2014-4}
  {\path{doi:10.12942/lrr-2014-4}}.

\bibitem{hawk}
S.~W. Hawking, G.~F.~R. Ellis, {The Large Scale Structure of Space-Time},
  Cambridge University Press, Cambridge, England, 1975.

\bibitem{wald}
R.~M. Wald, {General Relativity}, University of Chicago Press, Chacago and
  London, 1984.

\bibitem{copeland}
E.~J. {Copeland}, M.~{Sami}, S.~{Tsujikawa}, {Dynamics of Dark Energy},
  International Journal of Modern Physics D 15 (2006) 1753--1935.
\newblock \href {http://arxiv.org/abs/hep-th/0603057}
  {\path{arXiv:hep-th/0603057}}, \href
  {http://dx.doi.org/10.1142/S021827180600942X}
  {\path{doi:10.1142/S021827180600942X}}.

\bibitem{freese}
K.~{Freese}, {Status of Dark Matter in the Universe}, ArXiv
  e-prints~(1701.01840).
\newblock \href {http://arxiv.org/abs/1701.01840} {\path{arXiv:1701.01840}}.

\bibitem{mod_grav_rev2}
T.~Clifton, P.~G. Ferreira, A.~Padilla, C.~Skordis, Modified gravity and
  cosmology, Physics Reports 513~(1–3) (2012) 1 -- 189.
\newblock \href {http://dx.doi.org/10.1016/j.physrep.2012.01.001}
  {\path{doi:10.1016/j.physrep.2012.01.001}}.

\bibitem{mod_grav_rev1}
E.~Berti, E.~Barausse, V.~Cardoso, L.~Gualtieri, P.~Pani, U.~Sperhake, L.~C.
  Stein, N.~Wex, K.~Yagi, T.~Baker, et~al., {Testing general relativity with
  present and future astrophysical observations}, Classical and Quantum Gravity
  32~(24) (2015) 243001.

\bibitem{bojowald2001}
M.~Bojowald, {Absence of a Singularity in Loop Quantum Cosmology}, Phys. Rev.
  Lett. 86 (2001) 5227--5230.
\newblock \href {http://dx.doi.org/10.1103/PhysRevLett.86.5227}
  {\path{doi:10.1103/PhysRevLett.86.5227}}.

\bibitem{ashtekar2006}
A.~Ashtekar, T.~Pawlowski, P.~Singh, {Quantum Nature of the Big Bang}, Phys.
  Rev. Lett. 96 (2006) 141301.
\newblock \href {http://dx.doi.org/10.1103/PhysRevLett.96.141301}
  {\path{doi:10.1103/PhysRevLett.96.141301}}.

\bibitem{born}
M.~Born, L.~Infeld, {Foundations of the New Field Theory}, Proc. R. Soc. A
  144~(852) (1934) 425--451.

\bibitem{desgib}
S.~Deser, G.~W. Gibbons, {Born - Infeld - Einstein actions?}, Classical and
  Quantum Gravity 15~(5) (1998) L35.

\bibitem{edd}
A.~Eddington, {The Mathematical Theory of Relativity}, Cambridge University
  Press, Cambridge, England, 1924.

\bibitem{vollick}
D.~N. Vollick, {Palatini approach to Born-Infeld-Einstein theory and a
  geometric description of electrodynamics}, Phys. Rev. D 69 (2004) 064030.
\newblock \href {http://dx.doi.org/10.1103/PhysRevD.69.064030}
  {\path{doi:10.1103/PhysRevD.69.064030}}.

\bibitem{olmo2011}
G.~J. Olmo, {Palatini approach to modified gravity: f(R) theories and beyond},
  International Journal of Modern Physics D 20~(04) (2011) 413--462.
\newblock \href {http://dx.doi.org/10.1142/S0218271811018925}
  {\path{doi:10.1142/S0218271811018925}}.

\bibitem{vollick2}
D.~N. Vollick, {Born-Infeld-Einstein theory with matter}, Phys. Rev. D 72
  (2005) 084026.
\newblock \href {http://dx.doi.org/10.1103/PhysRevD.72.084026}
  {\path{doi:10.1103/PhysRevD.72.084026}}.

\bibitem{vollick3}
D.~N. {Vollick}, {Black hole and cosmological space-times in
  Born-Infeld-Einstein theory}, ArXiv e-prints~(gr-qc/0601136).
\newblock \href {http://arxiv.org/abs/gr-qc/0601136}
  {\path{arXiv:gr-qc/0601136}}.

\bibitem{banados}
M.~Ba\~nados, P.~G. Ferreira, {Eddington's Theory of Gravity and Its Progeny},
  Phys. Rev. Lett. 105 (2010) 011101.
\newblock \href {http://dx.doi.org/10.1103/PhysRevLett.105.011101}
  {\path{doi:10.1103/PhysRevLett.105.011101}}.

\bibitem{isham}
C.~J. Isham, A.~Salam, J.~Strathdee, {$f$-Dominance of Gravity}, Phys. Rev. D 3
  (1971) 867--873.
\newblock \href {http://dx.doi.org/10.1103/PhysRevD.3.867}
  {\path{doi:10.1103/PhysRevD.3.867}}.

\bibitem{scargill}
J.~H.~C. Scargill, M.~Banados, P.~G. Ferreira, {Cosmology with
  Eddington-inspired gravity}, Phys. Rev. D 86 (2012) 103533.
\newblock \href {http://dx.doi.org/10.1103/PhysRevD.86.103533}
  {\path{doi:10.1103/PhysRevD.86.103533}}.

\bibitem{schmidt}
A.~Schmidt-May, M.~von Strauss, {A link between ghost-free bimetric and
  Eddington-inspired Born-Infeld theory}\href {http://arxiv.org/abs/1412.3812}
  {\path{arXiv:1412.3812}}.

\bibitem{lavinia}
J.~B. Jiménez, L.~Heisenberg, G.~J. Olmo, {Infrared lessons for ultraviolet
  gravity: the case of massive gravity and Born-Infeld}, Journal of Cosmology
  and Astroparticle Physics 2014~(11) (2014) 004.

\bibitem{cardoso}
P.~Pani, V.~Cardoso, T.~Delsate, {Compact Stars in Eddington Inspired Gravity},
  Phys. Rev. Lett. 107 (2011) 031101.
\newblock \href {http://dx.doi.org/10.1103/PhysRevLett.107.031101}
  {\path{doi:10.1103/PhysRevLett.107.031101}}.

\bibitem{casanellas}
J.~Casanellas, P.~Pani, I.~Lopes, V.~Cardoso, {Testing Alternative Theories of
  Gravity Using the Sun}, The Astrophysical Journal 745~(1) (2012) 15.

\bibitem{avelino}
P.~P. Avelino, {Eddington-inspired Born-Infeld gravity: Astrophysical and
  cosmological constraints}, Phys. Rev. D 85 (2012) 104053.
\newblock \href {http://dx.doi.org/10.1103/PhysRevD.85.104053}
  {\path{doi:10.1103/PhysRevD.85.104053}}.

\bibitem{sham}
Y.-H. Sham, L.-M. Lin, P.~T. Leung, {Radial oscillations and stability of
  compact stars in Eddington-inspired Born-Infeld gravity}, Phys. Rev. D 86
  (2012) 064015.
\newblock \href {http://dx.doi.org/10.1103/PhysRevD.86.064015}
  {\path{doi:10.1103/PhysRevD.86.064015}}.

\bibitem{sham2}
Y.-H. Sham, P.~T. Leung, L.-M. Lin, {Compact stars in Eddington-inspired
  Born-Infeld gravity: Anomalies associated with phase transitions}, Phys. Rev.
  D 87 (2013) 061503.
\newblock \href {http://dx.doi.org/10.1103/PhysRevD.87.061503}
  {\path{doi:10.1103/PhysRevD.87.061503}}.

\bibitem{structure.exotic.star}
T.~Harko, F.~S.~N. Lobo, M.~K. Mak, S.~V. Sushkov, {Structure of neutron,
  quark, and exotic stars in Eddington-inspired Born-Infeld gravity}, Phys.
  Rev. D 88 (2013) 044032.
\newblock \href {http://dx.doi.org/10.1103/PhysRevD.88.044032}
  {\path{doi:10.1103/PhysRevD.88.044032}}.

\bibitem{sotani.neutron.star}
H.~Sotani, {Observational discrimination of Eddington-inspired Born-Infeld
  gravity from general relativity}, Phys. Rev. D 89 (2014) 104005.
\newblock \href {http://dx.doi.org/10.1103/PhysRevD.89.104005}
  {\path{doi:10.1103/PhysRevD.89.104005}}.

\bibitem{sotani.stellar.oscillations}
H.~Sotani, {Stellar oscillations in Eddington-inspired Born-Infeld gravity},
  Phys. Rev. D 89 (2014) 124037.
\newblock \href {http://dx.doi.org/10.1103/PhysRevD.89.124037}
  {\path{doi:10.1103/PhysRevD.89.124037}}.

\bibitem{sotani.magnetic.star}
H.~Sotani, {Magnetized relativistic stellar models in Eddington-inspired
  Born-Infeld gravity}, ArXiv e-prints\href {http://arxiv.org/abs/1503.07942}
  {\path{arXiv:1503.07942}}.

\bibitem{wei}
S.-W. {Wei}, K.~{Yang}, Y.-X. {Liu}, {Black hole solution and strong
  gravitational lensing in Eddington-inspired Born-Infeld gravity}, European
  Physical Journal C 75 (2015) 253.
\newblock \href {http://arxiv.org/abs/1405.2178} {\path{arXiv:1405.2178}},
  \href {http://dx.doi.org/10.1140/epjc/s10052-015-3469-7}
  {\path{doi:10.1140/epjc/s10052-015-3469-7}}.

\bibitem{sotani}
H.~Sotani, U.~Miyamoto, {Properties of an electrically charged black hole in
  Eddington-inspired Born-Infeld gravity}, Phys. Rev. D 90 (2014) 124087.
\newblock \href {http://dx.doi.org/10.1103/PhysRevD.90.124087}
  {\path{doi:10.1103/PhysRevD.90.124087}}.

\bibitem{eibiwormhole}
G.~J. Olmo, D.~Rubiera-Garcia, H.~Sanchis-Alepuz, {Geonic black holes and
  remnants in Eddington-inspired Born-Infeld gravity}, The European Physical
  Journal C 74~(3) (2014) 2804.
\newblock \href {http://dx.doi.org/10.1140/epjc/s10052-014-2804-8}
  {\path{doi:10.1140/epjc/s10052-014-2804-8}}.

\bibitem{rajibul}
R.~Shaikh, {Lorentzian wormholes in Eddington-inspired Born-Infeld gravity},
  Phys. Rev. D 92 (2015) 024015.
\newblock \href {http://dx.doi.org/10.1103/PhysRevD.92.024015}
  {\path{doi:10.1103/PhysRevD.92.024015}}.

\bibitem{jana2}
S.~Jana, S.~Kar, {Born-Infeld gravity coupled to Born-Infeld electrodynamics},
  Phys. Rev. D 92 (2015) 084004.
\newblock \href {http://dx.doi.org/10.1103/PhysRevD.92.084004}
  {\path{doi:10.1103/PhysRevD.92.084004}}.

\bibitem{BTZ_typesoln}
D.~Bazeia, L.~Losano, G.~J. Olmo, D.~Rubiera-Garcia, {Geodesically complete
  BTZ-type solutions of $2  +  1$ Born-Infeld gravity}, Classical and
  Quantum Gravity 34~(4) (2017) 045006.

\bibitem{scalar_geon_BIgravity}
V.~I. {Afonso}, G.~J. {Olmo}, D.~{Rubiera-Garcia}, {Scalar geons in Born-Infeld
  gravity}, ArXiv e-prints\href {http://arxiv.org/abs/1705.01065}
  {\path{arXiv:1705.01065}}.

\bibitem{eibibrane}
Y.-X. Liu, K.~Yang, H.~Guo, Y.~Zhong, {Domain wall brane in Eddington-inspired
  Born-Infeld gravity}, Phys. Rev. D 85 (2012) 124053.
\newblock \href {http://dx.doi.org/10.1103/PhysRevD.85.124053}
  {\path{doi:10.1103/PhysRevD.85.124053}}.

\bibitem{delsate}
T.~Delsate, J.~Steinhoff, {New Insights on the Matter-Gravity Coupling
  Paradigm}, Phys. Rev. Lett. 109 (2012) 021101.
\newblock \href {http://dx.doi.org/10.1103/PhysRevLett.109.021101}
  {\path{doi:10.1103/PhysRevLett.109.021101}}.

\bibitem{cho_prd88}
I.~Cho, H.-C. Kim, {New synthesis of matter and gravity: A nongravitating
  scalar field}, Phys. Rev. D 88 (2013) 064038.
\newblock \href {http://dx.doi.org/10.1103/PhysRevD.88.064038}
  {\path{doi:10.1103/PhysRevD.88.064038}}.

\bibitem{jana}
S.~Jana, S.~Kar, {Three dimensional Eddington-inspired Born-Infeld gravity:
  Solutions}, Phys. Rev. D 88 (2013) 024013.
\newblock \href {http://dx.doi.org/10.1103/PhysRevD.88.024013}
  {\path{doi:10.1103/PhysRevD.88.024013}}.

\bibitem{pani}
P.~Pani, T.~P. Sotiriou, {Surface Singularities in Eddington-Inspired
  Born-Infeld Gravity}, Phys. Rev. Lett. 109 (2012) 251102.
\newblock \href {http://dx.doi.org/10.1103/PhysRevLett.109.251102}
  {\path{doi:10.1103/PhysRevLett.109.251102}}.

\bibitem{eibiprob.cure}
H.-C. Kim, {Physics at the surface of a star in Eddington-inspired Born-Infeld
  Gravity}, Phys. Rev. D 89 (2014) 064001.
\newblock \href {http://dx.doi.org/10.1103/PhysRevD.89.064001}
  {\path{doi:10.1103/PhysRevD.89.064001}}.

\bibitem{odintsov}
S.~D. Odintsov, G.~J. Olmo, D.~Rubiera-Garcia, {Born-Infeld gravity and its
  functional extensions}, Phys. Rev. D 90 (2014) 044003.
\newblock \href {http://dx.doi.org/10.1103/PhysRevD.90.044003}
  {\path{doi:10.1103/PhysRevD.90.044003}}.

\bibitem{fernandes}
K.~Fernandes, A.~Lahiri, {Kaluza Ansatz applied to Eddington inspired
  Born-Infeld gravity}, Phys. Rev. D 91 (2015) 044014.
\newblock \href {http://dx.doi.org/10.1103/PhysRevD.91.044014}
  {\path{doi:10.1103/PhysRevD.91.044014}}.

\bibitem{chen2016}
C.-Y. {Chen}, M.~{Bouhmadi-L{\'o}pez}, P.~{Chen}, {Modified
  Eddington-inspired-Born-Infeld Gravity with a Trace Term}, European Physical
  Journal C 76 (2016) 40.
\newblock \href {http://arxiv.org/abs/1507.00028} {\path{arXiv:1507.00028}},
  \href {http://dx.doi.org/10.1140/epjc/s10052-016-3879-1}
  {\path{doi:10.1140/epjc/s10052-016-3879-1}}.

\bibitem{cho}
I.~Cho, H.-C. Kim, T.~Moon, {Universe driven by a perfect fluid in
  Eddington-inspired Born-Infeld gravity}, Phys. Rev. D 86 (2012) 084018.
\newblock \href {http://dx.doi.org/10.1103/PhysRevD.86.084018}
  {\path{doi:10.1103/PhysRevD.86.084018}}.

\bibitem{escamilla}
C.~Escamilla-Rivera, M.~Banados, P.~G. Ferreira, {Tensor instability in the
  Eddington-inspired Born-Infeld theory of gravity}, Phys. Rev. D 85 (2012)
  087302.
\newblock \href {http://dx.doi.org/10.1103/PhysRevD.85.087302}
  {\path{doi:10.1103/PhysRevD.85.087302}}.

\bibitem{linear.perturbation}
K.~Yang, X.-L. Du, Y.-X. Liu, {Linear perturbations in Eddington-inspired
  Born-Infeld gravity}, Phys. Rev. D 88 (2013) 124037.
\newblock \href {http://dx.doi.org/10.1103/PhysRevD.88.124037}
  {\path{doi:10.1103/PhysRevD.88.124037}}.

\bibitem{avelinoferreira}
P.~P. Avelino, R.~Z. Ferreira, {Bouncing Eddington-inspired Born-Infeld
  cosmologies: An alternative to inflation?}, Phys. Rev. D 86 (2012) 041501.
\newblock \href {http://dx.doi.org/10.1103/PhysRevD.86.041501}
  {\path{doi:10.1103/PhysRevD.86.041501}}.

\bibitem{chokim}
I.~Cho, H.-C. Kim, T.~Moon, {Precursor of Inflation}, Phys. Rev. Lett. 111
  (2013) 071301.
\newblock \href {http://dx.doi.org/10.1103/PhysRevLett.111.071301}
  {\path{doi:10.1103/PhysRevLett.111.071301}}.

\bibitem{cho90}
I.~Cho, H.-C. Kim, {Inflationary tensor perturbation in Eddington-inspired
  Born-Infeld gravity}, Phys. Rev. D 90 (2014) 024063.
\newblock \href {http://dx.doi.org/10.1103/PhysRevD.90.024063}
  {\path{doi:10.1103/PhysRevD.90.024063}}.

\bibitem{cho_scalar_perturbation}
I.~Cho, N.~K. Singh, {Scalar perturbation produced at the pre-inflationary
  stage in Eddington-inspired Born-Infeld gravity}, European Physical Journal C
  75 (2015) 240.
\newblock \href {http://arxiv.org/abs/1412.6344} {\path{arXiv:1412.6344}},
  \href {http://dx.doi.org/10.1140/epjc/s10052-015-3458-x}
  {\path{doi:10.1140/epjc/s10052-015-3458-x}}.

\bibitem{large.scale.structure}
X.-L. Du, K.~Yang, X.-H. Meng, Y.-X. Liu, {Large scale structure formation in
  Eddington-inspired Born-Infeld gravity}, Phys. Rev. D 90 (2014) 044054.
\newblock \href {http://dx.doi.org/10.1103/PhysRevD.90.044054}
  {\path{doi:10.1103/PhysRevD.90.044054}}.

\bibitem{Bouhmadi-Lopez2016}
M.~Bouhmadi-L{\' o}pez, C.-Y. Chen, {Towards the quantization of
  Eddington-inspired-Born-Infeld theory}, Journal of Cosmology and
  Astroparticle Physics 2016~(11) (2016) 023.

\bibitem{lopez2017}
I.~{Albarran}, M.~{Bouhmadi-L{\'o}pez}, C.-Y. {Chen}, P.~{Chen}, {Doomsdays in
  a modified theory of gravity: A classical and a quantum approach}, ArXiv
  e-prints\href {http://arxiv.org/abs/1703.09263} {\path{arXiv:1703.09263}}.

\bibitem{cho_spectral_indices}
I.~Cho, J.-O. Gong, {Spectral indices in Eddington-inspired Born-Infeld
  inflation}, Phys. Rev. D 92 (2015) 064046.
\newblock \href {http://dx.doi.org/10.1103/PhysRevD.92.064046}
  {\path{doi:10.1103/PhysRevD.92.064046}}.

\bibitem{cho_pow_spectra}
I.~Cho, N.~K. Singh, {Primordial power spectra of Eddington-inspired
  Born-Infeld inflation in strong gravity limit}, Phys. Rev. D 92 (2015)
  024038.
\newblock \href {http://dx.doi.org/10.1103/PhysRevD.92.024038}
  {\path{doi:10.1103/PhysRevD.92.024038}}.

\bibitem{cho_tensor_scalar}
I.~{Cho}, N.~K. {Singh}, {Tensor-to-scalar ratio in Eddington-inspired
  Born-Infeld inflation}, European Physical Journal C 74 (2014) 3155.
\newblock \href {http://arxiv.org/abs/1408.2652} {\path{arXiv:1408.2652}},
  \href {http://dx.doi.org/10.1140/epjc/s10052-014-3155-1}
  {\path{doi:10.1140/epjc/s10052-014-3155-1}}.

\bibitem{felice}
A.~De~Felice, B.~Gumjudpai, S.~Jhingan, {Cosmological constraints for an
  Eddington-Born-Infeld field}, Phys. Rev. D 86 (2012) 043525.
\newblock \href {http://dx.doi.org/10.1103/PhysRevD.86.043525}
  {\path{doi:10.1103/PhysRevD.86.043525}}.

\bibitem{power.spectrum}
M.~Lagos, M.~Ba\~nados, P.~G. Ferreira, S.~Garcia-Saenz, {Noether identities
  and gauge fixing the action for cosmological perturbations}, Phys. Rev. D 89
  (2014) 024034.
\newblock \href {http://dx.doi.org/10.1103/PhysRevD.89.024034}
  {\path{doi:10.1103/PhysRevD.89.024034}}.

\bibitem{cascading_dust_inflation}
J.~B. Jiménez, L.~Heisenberg, G.~J. Olmo, C.~Ringeval, {Cascading dust
  inflation in Born-Infeld gravity}, Journal of Cosmology and Astroparticle
  Physics 2015~(11) (2015) 046.

\bibitem{bianchi.cosmo}
T.~Harko, F.~S. Lobo, M.~K. Mak, {Bianchi Type I Cosmological Models in
  Eddington-inspired Born–Infeld Gravity}, Galaxies 2~(4) (2014) 496--519.
\newblock \href {http://dx.doi.org/10.3390/galaxies2040496}
  {\path{doi:10.3390/galaxies2040496}}.

\bibitem{jana4}
S.~Jana, S.~Kar, {Born-Infeld cosmology with scalar Born-Infeld matter}, Phys.
  Rev. D 94 (2016) 064016.
\newblock \href {http://dx.doi.org/10.1103/PhysRevD.94.064016}
  {\path{doi:10.1103/PhysRevD.94.064016}}.

\bibitem{instanton}
F.~Arroja, C.-Y. Chen, P.~Chen, D.~han Yeom, {Singular instantons in
  Eddington-inspired-Born-Infeld gravity}, Journal of Cosmology and
  Astroparticle Physics 2017~(03) (2017) 044.

\bibitem{Bouhmadi-Lopez2014}
M.~Bouhmadi-L{\'o}pez, C.-Y. Chen, P.~Chen, {Is Eddington--Born--Infeld theory
  really free of cosmological singularities?}, The European Physical Journal C
  74~(3) (2014) 2802.
\newblock \href {http://dx.doi.org/10.1140/epjc/s10052-014-2802-x}
  {\path{doi:10.1140/epjc/s10052-014-2802-x}}.

\bibitem{Bouhmadi-Lopez2015}
M.~Bouhmadi-L{\'o}pez, C.-Y. Chen, P.~Chen, {Eddington--Born--Infeld cosmology:
  a cosmographic approach, a tale of doomsdays and the fate of bound
  structures}, The European Physical Journal C 75~(2) (2015) 90.
\newblock \href {http://dx.doi.org/10.1140/epjc/s10052-015-3257-4}
  {\path{doi:10.1140/epjc/s10052-015-3257-4}}.

\bibitem{lopez2014}
M.~{Bouhmadi-L{\'o}pez}, C.-Y. {Chen}, P.~{Chen}, {Cosmological singularities
  in Born-Infeld determinantal gravity}, Phys. Rev. D 90~(1407.5114) (2014)
  123518.
\newblock \href {http://arxiv.org/abs/1407.5114} {\path{arXiv:1407.5114}},
  \href {http://dx.doi.org/10.1103/PhysRevD.90.123518}
  {\path{doi:10.1103/PhysRevD.90.123518}}.

\bibitem{jimenez2017}
J.~B. {Jimenez}, L.~{Heisenberg}, G.~J. {Olmo}, D.~{Rubiera-Garcia},
  {Born-Infeld inspired modifications of gravity}, ArXiv e-prints~(1704.03351).
\newblock \href {http://arxiv.org/abs/1704.03351} {\path{arXiv:1704.03351}}.

\bibitem{nuclear.test}
P.~Avelino, {Eddington-inspired Born-Infeld gravity: nuclear physics
  constraints and the validity of the continuous fluid approximation}, Journal
  of Cosmology and Astroparticle Physics 2012~(11) (2012) 022.

\bibitem{faraoni}
V.~Faraoni, {Cosmology in Scalar-Tensor Gravity}, Springer, New York, 2004.

\bibitem{nojiri2005}
S.~Nojiri, S.~D. Odintsov, S.~Tsujikawa, {Properties of singularities in the
  (phantom) dark energy universe}, Phys. Rev. D 71 (2005) 063004.
\newblock \href {http://dx.doi.org/10.1103/PhysRevD.71.063004}
  {\path{doi:10.1103/PhysRevD.71.063004}}.

\bibitem{nojiri2005b}
S.~Nojiri, S.~D. Odintsov, Inhomogeneous equation of state of the universe:
  Phantom era, future singularity, and crossing the phantom barrier, Phys. Rev.
  D 72 (2005) 023003.
\newblock \href {http://dx.doi.org/10.1103/PhysRevD.72.023003}
  {\path{doi:10.1103/PhysRevD.72.023003}}.

\bibitem{jimenez2016}
J.~B. Jim\'enez, D.~Rubiera-Garcia, D.~S\'aez-G\'omez, V.~Salzano,
  {Cosmological future singularities in interacting dark energy models}, Phys.
  Rev. D 94 (2016) 123520.
\newblock \href {http://dx.doi.org/10.1103/PhysRevD.94.123520}
  {\path{doi:10.1103/PhysRevD.94.123520}}.

\end{thebibliography}
\end{document}